\documentclass[aps,prl,twocolumn,footinbib,showpacs,preprintnumbers,notitlepage,superscriptaddress,longbibliography]{revtex4-2}
    \usepackage{graphicx}
\usepackage{graphics}
\usepackage{amsmath}
\usepackage{amssymb}
\usepackage{amsfonts}
\usepackage{dsfont}
\usepackage{braket}
\usepackage{color}
\usepackage{braket,slashed}
\usepackage[mathscr]{euscript}
\definecolor{darkblue}{rgb}{0, 0, 0.8}
\usepackage[colorlinks=true, breaklinks=true, linkcolor=red, citecolor=blue, urlcolor=blue]{hyperref} 
\usepackage{subfigure}
\usepackage{xfrac}
\usepackage{bm}
\usepackage{kantlipsum}
\usepackage{enumitem}
\usepackage{tikz}
\usepackage{framed}
\usepackage{graphicx}
\usepackage{subfigure}
\usepackage{cleveref}
\usepackage{adjustbox}

\usepackage{xcolor}

\allowdisplaybreaks[1]
    
    \usepackage[english]{babel}
\begin{document}
    
    \title{Incommensurate Order with Translationally Invariant Projected Entangled-Pair States:
    Spiral States and Quantum Spin Liquid on the Anisotropic Triangular Lattice}
    
    \author{Juraj Hasik}
    \affiliation{Institute for Theoretical Physics, University of Amsterdam,
    Science Park 904, 1098 XH Amsterdam, The Netherlands}
    
    \author{Philippe Corboz}
    \affiliation{Institute for Theoretical Physics, University of Amsterdam,
    Science Park 904, 1098 XH Amsterdam, The Netherlands}
    
    \begin{abstract}
    Simulating strongly correlated systems with incommensurate order poses significant challenges for traditional finite-size-based approaches. 
    Confining such a phase to a finite-size geometry can induce spurious frustration, with spin spirals in frustrated magnets being a typical example.
    Here, we introduce an Ansatz based on infinite projected entangled-pair states (iPEPS) which overcomes these limitations and enables the direct search for the optimal spiral in the thermodynamic limit, with a computational cost that is independent of the spiral's wavelength.  
    Leveraging this method, we simulate the Heisenberg model on the anisotropic triangular lattice, which interpolates between the square and isotropic triangular lattice limits. Besides accurately reproducing the magnetically ordered phases with arbitrary wavelength, the simulations reveal a quantum spin liquid phase emerging between the N\'eel and spin spiral phases.
    \end{abstract}
    
    \maketitle
    
    The ongoing quest for a quantitative treatment of strongly correlated condensed matter systems 
    is constantly moving forward, stimulated by challenges such as modeling of high-$T_c$ superconductivity or searching for topological phases of matter, with quantum spin liquids being a prime example~\cite{Wen1989,balents2010}.
    Within exact and variational approaches, an extremely successful paradigm has been the simulation of finite-size systems by exact diagonalization, quantum and variational Monte Carlo (VMC), and density-matrix renormalization group (DMRG)~\cite{white1992}.
    More recently, the modern formulation of DMRG within the tensor network framework led to a generalization of the underlying matrix product state (MPS)  Ansatz~\cite{ostlund1995,perez2007} to two dimensions, dubbed projected entangled-pair states (PEPS)~\cite{Verstraete2004,nishio2004}. Both MPS and PEPS Ansätze can be consistently formulated directly in the thermodynamic limit~\cite{Jordan2008}, where a unit cell containing all nonequivalent tensors tiles the infinite lattice, presenting an alternative paradigm for simulations of strongly correlated systems. 
    Capabilities of infinite PEPS (iPEPS) were demonstrated in numerous works addressing the phase diagram of challenging models, such as   doped Hubbard models~\cite{zheng17,ponsioen19,zhang23b,ponsioen2023superconducting} or frustrated antiferromagnets  the square~\cite{niesen17,chen18,hasik21,liu22b,hasik22}, Kagome~\cite{PhysRevLett.118.137202,kshetrimayum19b}, and other lattices~\cite{PhysRevB.87.115144,corboz14_shastry,lee18,jahromi18,lee20,zhang23}.

    Among various conventional orders,  incommensurate phases such as spin spirals pose a challenge to both the finite-size approaches and iPEPS. Spin spirals are created by local magnetic moments forming a periodic pattern with wavelengths that can be much larger than the lattice spacing.
    Typically, finite-size systems are treated together with periodic boundary conditions, which results in spurious frustration when such phase is confined to a simulation cell that is incommensurate with the wavelength. Open boundaries alleviate such frustration at the cost of larger finite-size effects.  
    Simulating such phases with iPEPS was deemed difficult if not impossible, due to an apparent need for unit cells large enough to accommodate such spirals, making the simulation prohibitively expensive. 
    
    Here, we extend iPEPS by introducing a unitary $U(\mathbf{q})$ parametrized by a wave vector $\mathbf{q}$ which can be variationally optimized 
    together with the tensors of the network. 
    We employ this Ansatz, which we term spiral iPEPS, to study the incommensurate phase appearing in the anisotropic Heisenberg model on the triangular lattice, investigated previously by perturbative approaches~\cite{PhysRevB.59.14367,gutzwiller2007powell,ccluster2009bishop,spinwave2011hauke,schwinger2020gonzalez}, functional renormalization group~\cite{reuther11}, and variational methods ~\cite{dmrg2011weichselbaum, PhysRevB.93.085111}. 
    We show how a spiral iPEPS faithfully captures these phases with optimized $U(\mathbf{q})$, creating the correct magnetic texture, while the underlying iPEPS, generated by only a single tensor, accounts for quantum-mechanical correlations. Crucially, the computational complexity remains independent of the wave vector $\mathbf{q}$ and scales only with the size of the tensors.
    In one dimension, a similar approach has been applied to the MPS in Ref.~\cite{ueda2012}. 
    
   Besides magnetically ordered phases, we find strong evidence of a quantum spin liquid (QSL) emerging in between,
   as revealed by a systematic scaling analysis of the order parameter. A QSL was also recently proposed in a VMC study~\cite{PhysRevB.93.085111}, which, however, was limited to only a few discrete values of $\mathbf{q}$ due to the finite system size. With spiral iPEPS, in contrast, we can probe any value of $\mathbf{q}$ in order to rule out that the QSL is an artifact of spurious frustration. 
A spiral iPEPS thus complements a standard iPEPS with multisite unit cells and together enable a more systematic and unbiased exploration of phase diagrams.
   This makes it an ideal approach to study anisotropic triangular lattice compounds that may host a quantum spin liquid phase, such as certain organic Pd(dmit)$_2$  and $\kappa$-(ET)$_2X$ compounds~\cite{powell11,kanoda11,shimizu03, manna10}, or spin-spiral order as found in Cs$_2$CuCl$_4$~\cite{coldea03}.
    
    \begin{figure}[tb]
        \includegraphics[width=\columnwidth]{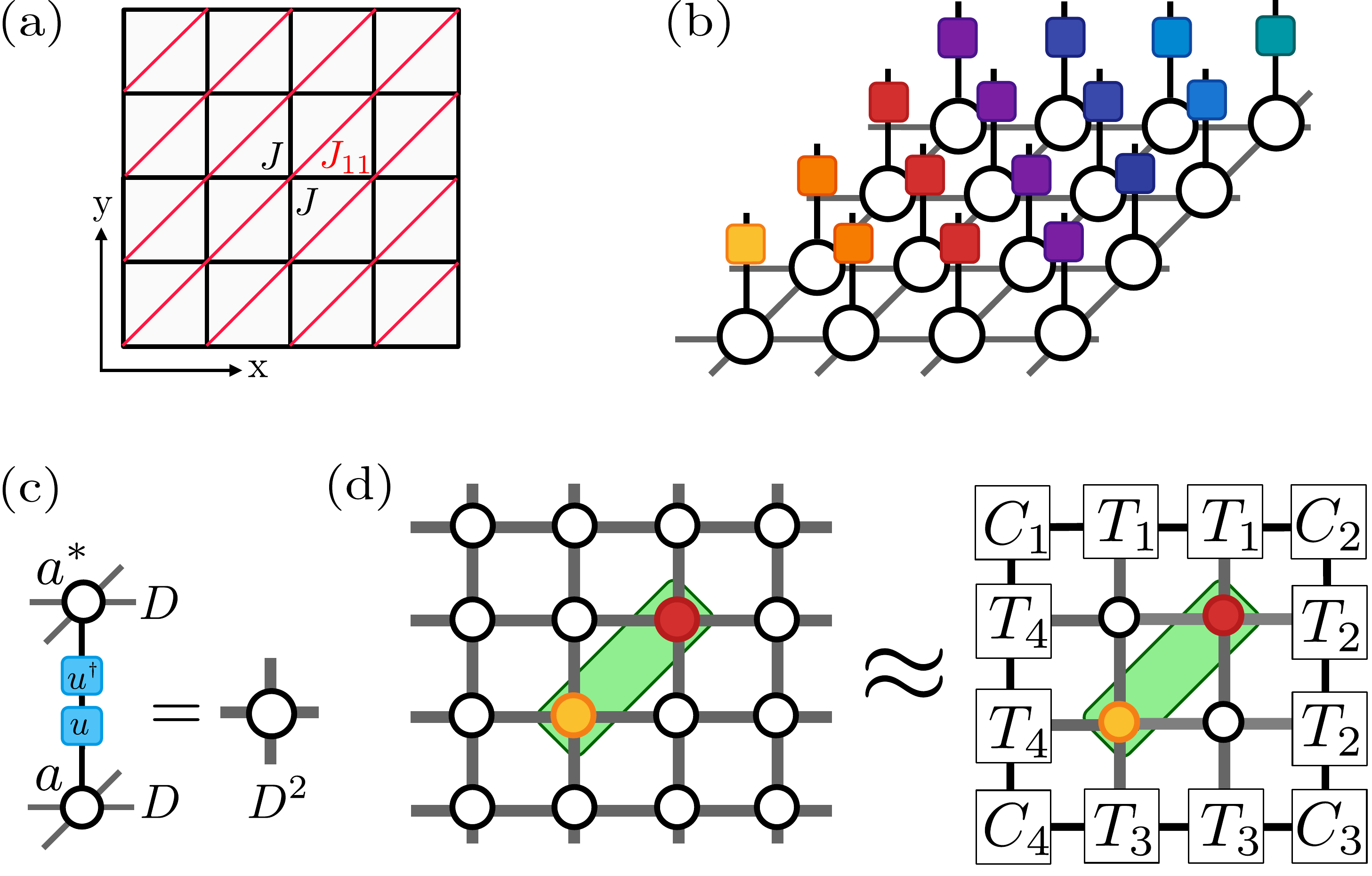}
        \caption{(a) The anisotropic triangular Heisenberg model (Eq.~\ref{eq:model}), here represented with couplings $J$ on a square lattice and $J_{11}$ along one of the  diagonal directions.
        (b) Spiral iPEPS Ansatz $|\psi(a,\mathbf{q})\rangle$, where vertical black lines denote physical indices and gray lines in plane denote auxiliary indices of the on-site tensors (circles) with bond dimension $D$. The boxes represent position-dependent unitaries $u_{\mathbf{r}}(\mathbf{q},\mathbf{r})$, with color encoding the phase. (c) Definition of the double-layer tensor. Note that the local unitaries cancel. (d)  Expectation value of the next-nearest-neighbor exchange $\langle\mathbf{S}_\mathbf{r}\cdot\mathbf{S}_{\mathbf{r}+\hat{x}+\hat{y}}\rangle$ (the exchange operator is denoted by a green box), computed using the  CTMRG environment, with thick black lines representing indices of dimension $\chi$ and thick gray lines double-layer auxiliary indices of dimension $D^2$. The colored circles represent on-site tensors together with remaining local unitaries $u_{\mathbf{r}}(\mathbf{q},\mathbf{r})$.}
        \label{fig:lattice} 
    \end{figure}
    
    
    \emph{{The model.}}--- The antiferromagnetic Heisenberg model on the anisotropic triangular lattice is described by the SU(2)-symmetric spin-$1/2$ Hamiltonian,
    \begin{equation}
        H = \sum_{\mathbf{r}} J(\mathbf{S_r}\cdot\mathbf{S}_{\mathbf{r}+\hat{x}} + \mathbf{S_r}\cdot\mathbf{S}_{\mathbf{r}+\hat{y}}) + J_{11} \mathbf{S}_\mathbf{r}\cdot\mathbf{S}_{\mathbf{r}+\hat{x}+\hat{y}},
        \label{eq:model}
    \end{equation}
    where $\mathbf{S}_r = (S^x, S^y, S^z)$ denotes the vector of spin-$1/2$ operators at site $\mathbf{r}$, and $\hat{x}$, $\hat{y}$ are unit vectors in horizontal and vertical directions of the lattice, respectively.
    We set $J$ to unity and focus on the interval of diagonal exchange coupling $J_{11} \in [0,1]$.
    In the limit of $J_{11}=0$ and $J_{11}=1$ the model reduces to the Heisenberg antiferromagnet on the square lattice and the isotropic triangular lattice, respectively, see Fig. \hyperref[fig:lattice]{1(a)}. The corresponding ground states spontaneously break the SU(2) spin symmetry, where the first case displays a collinear $\mathbf{q}=(\pi,\pi)$ order, characteristic of the N\'eel phase, while the latter orders with $\mathbf{q}=(\pm 2\pi/3, \pm 2\pi/3)$ in plane with $120^\circ$ angle between the spins on three different sublattices~\cite{PhysRevLett.82.3899,white2007order}.
    The strong frustration in the system at intermediate anisotropies can destabilize the magnetic order.
    Such instability was observed for $J_{11}\in[0.77,0.88]$ by spin-wave theory~\cite{spinwave2011hauke},
    or, for $J_{11}\in[0.6,0.9]$ by Schwinger boson theory~\cite{schwinger2020gonzalez}, which suggested a zero flux or a nematic QSL as candidate ground states. The existence and nature of this paramagnetic phase is debated. 
    VMC simulations~\cite{PhysRevB.93.085111} revealed the close competition of a gapless QSL and spiral states within $J_{11}\in[0.7,0.8]$, which, however, was found hard to accurately resolve  for the largest $L=18$ system considered. Alternatively, 
    a dimerized state was proposed for $J_{11}\in[0.7, 0.9]$ by a series expansion (SE) study~\cite{PhysRevB.59.14367}. Weak spatial symmetry breaking was also observed in this region by DMRG on finite-width cylinders~\cite{dmrg2011weichselbaum}, yet its  fate could not be conclusively answered due to finite size effects.

    \emph{{Spiral iPEPS ansatz and observables.}}--- 
    An iPEPS encodes the coefficients of a wave function by a set $\{a,b,c,\ldots\}$ of rank-five tensors, each with a physical index of dimension two corresponding to spin-1/2 degree of freedom and four auxiliary indices of bond dimension $D$, controlling the accuracy of the Ansatz. 
    These tensors, arranged in a unit cell which then tiles the infinite square lattice, are contracted by pairs of adjacent auxiliary indices. A standard example of an iPEPS designed to capture antiferromagnetic  $\mathbf{q}=(\pi,\pi)$ order uses two tensors $a$, $b$ arranged in a bipartite pattern, as illustrated below. An alternative approach to study antiferromagnetic Heisenberg models involves first transforming the Hamiltonian by a local unitary transformation, $u=-2iS^y$, on each $b$-sublattice site~\cite{10.21468/SciPostPhys.10.1.012,PhysRevB.96.121118}. This transformation  flips the spins as $u\lvert\uparrow\rangle = \lvert\downarrow\rangle$ and $u\lvert\downarrow\rangle = -\lvert\uparrow\rangle$.  The ground state of the transformed Hamiltonian then exhibits ferromagnetic order, which can be captured by a single-tensor Ansatz.
    Equivalently, instead of transforming the Hamiltonian, one can use the original Hamiltonian in conjunction with a single-tensor Ansatz that includes the unitaries $u$ on the $b$-sublattice to represent the antiferromagnetic ground state, 
    \begin{equation}
        \sum_{\{\mathbf{s}\}}\vcenter{\hbox{\includegraphics[scale=0.13]{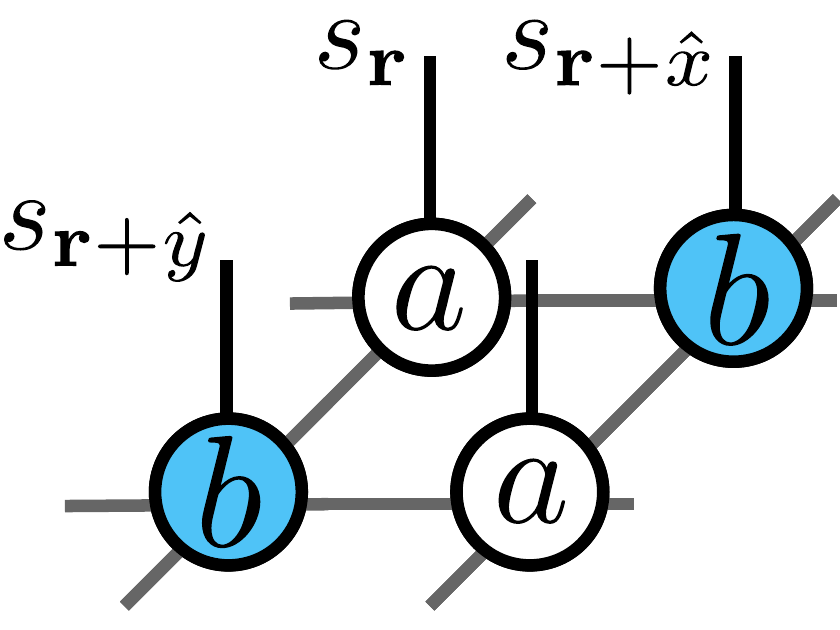}}}|\mathbf{s}\rangle=
        \sum_{\{\mathbf{s}\}}\vcenter{\hbox{\includegraphics[scale=0.13]{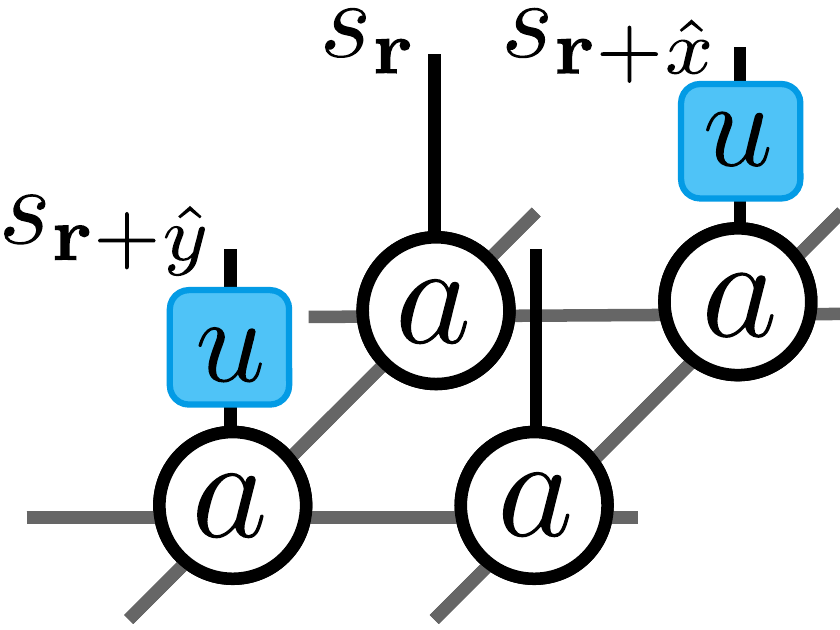}}}|\mathbf{s}\rangle
        =:U|\textrm{iPEPS(a)}\rangle,
    \label{eq:neelipeps}
    \end{equation}
    where circles denote tensors and in-plane lines correspond to  the auxiliary indices. The latter formulation will be advantageous for the generalization that follows.

    For the magnetically ordered phases of the model in Eq.~\ref{eq:model},
    the local magnetic moments lie in plane and their magnitude $m = |\langle \mathbf{S}_\mathbf{r} \rangle|$ is uniform as the ordering vector $\mathbf{q}$ changes when $J_{11}$ is varied.
    This observation motivates us to generalize the Ansatz of Eq.~\ref{eq:neelipeps} and propose a spiral iPEPS,
    \begin{equation}\label{eq:ansatz}
    |\psi(a,\mathbf{q})\rangle = U(\mathbf{q})|\textrm{iPEPS}(a)\rangle,
    \end{equation} 
    with a unitary $U(\mathbf{q})$ parametrized by a wave vector $\mathbf{q}$, shown in Fig.~\hyperref[fig:lattice]{1(b)}. Importantly,
    this unitary is a product of spatially dependent but local unitaries  $U(\mathbf{q})=\prod_\mathbf{r} u_\mathbf{r}(\mathbf{q},\mathbf{r})$ with $u_\mathbf{r}(\mathbf{q},\mathbf{r'}) = \exp[ i (\mathbf{q}\cdot\mathbf{r'}) S^y_{\mathbf{r}}]$. Intuitively, the $|\textrm{iPEPS}(a)\rangle$ will generate correlations which are independent of the position, while the  unitary $U(\mathbf{q})$ helps to realize optimal magnetic texture in the $xz$ plane. Optimization of both the on-site tensor $a$ and the wave vector $\mathbf{q}$ opens a new path, enabling us to search directly in the variational space of spiral states. 
    
    To evaluate observables $\mathcal{O}$, in particular the energy per site $e$, the local magnetic moment $m$, or the correlation length $\xi$ of a state, we employ the corner transfer matrix renormalization group (CTMRG)~\cite{nishino1996,orus2009-1,Corboz2014} method. It approximates formally infinitely large parts of the double-layer tensor network representing the expectation value of a local observable, $\langle \mathcal{O} \rangle = \langle \psi(a,\mathbf{q})| \mathcal{O} |\psi(a,\mathbf{q})\rangle$, by a set of eight environment tensors: 
    four corners $\{C_1,\ldots,C_4\}$ and four half-row and half-column tensors $\{T_1,\ldots,T_4\}$ with finite environment dimension $\chi$ controlling the precision of this approximation. 
    In Figs.~\hyperref[fig:lattice]{\ref*{fig:lattice}(c)} and ~\hyperref[fig:lattice]{\ref*{fig:lattice}(d)} we illustrate the CTMRG environment for $\langle\mathbf{S}_\mathbf{r}\cdot\mathbf{S}_{\mathbf{r}+\hat{x}+\hat{y}}\rangle$.

    The structure of the Ansatz together with the translational and the SU(2) symmetry of the Hamiltonian greatly simplify the evaluation of observables. First, the local unitaries $u_\mathbf{r}(\mathbf{q},\mathbf{r})$ cancel out in the double-layer tensor network (except where the observable is located), thus leaving the environment tensors independent of the wave vector $\mathbf{q}$. Hence, even if the physical unit cell accommodating a $\mathbf{q}$ spiral spans many lattice sites we do not need new environment tensors beyond the original eight $\{C_1,\ldots,C_4,T_1,\ldots,T_4\}$ and 
    environment of any rectangular region, where the observable $\mathcal{O}$ is supported, can be obtained just from these tensors.
    A complementary interpretation of the Ansatz gives an insight into the second simplification. For any given $J_{11}$ we find the optimal wave vector $\mathbf{q}$, which renders the ground state of the transformed
    Hamiltonian $H(\mathbf{q})= U^{\dag} (\mathbf{q})HU(\mathbf{q})$ to be a translationally invariant correlated ferromagnet,
    and then search for its variational description in terms of a single-site $|\textrm{iPEPS}(a)\rangle$.
    Because of the SU(2)-symmetry of the Hamiltonian the action of $U(\mathbf{q})$ on the individual exchange terms
    depends only on the relative position of the spins involved,
    $U^{\dag}(\mathbf{q})\mathbf{S}_\mathbf{r}\cdot\mathbf{S}_{\mathbf{r'}}U(\mathbf{q}) = \mathbf{S}_\mathbf{r}\cdot\mathbf{\tilde{S}}_{\mathbf{r'}}(\mathbf{q},\Delta)$, 
    where $\tilde{S}^\alpha_{\mathbf{r'}}(\mathbf{q},\Delta) = u^{\dag}_\mathbf{r'}(\mathbf{q},\Delta)S^\alpha u_\mathbf{r'}(\mathbf{q},\Delta)$ with $\Delta=\mathbf{r-r'}$. This leaves the energy per site $e$ of the transformed Hamiltonian uniform and composed only from three nonequivalent nearest-neighbor spin exchange terms,
    which we evaluate using the CTMRG environment. 
    
    \emph{{Finite-$D$ results.}}---
    We analyze the phase diagram of the model in Eq.~\ref{eq:model} by performing a variational optimization of the spiral iPEPS Ansatz~(\ref{eq:ansatz}). For each bond dimension $D$ and increasingly larger environment dimension $\chi$, we optimize states using gradient descent.
    The gradients of the energy $e$ are evaluated by automatic differentiation~\cite{Liao2017a} with respect to both real-valued on-site tensor $a$ and the wave vector parametrized by a single real-valued $q$ as $\mathbf{q}=(2\pi q,2\pi q)$, i.e. in total $2 D^4 + 1$ variational parameters. For further details regarding implementation and optimizations of spiral iPEPS, see (SM)~\cite{SM}. 
    For the largest $D=6$ considered here we reach $\chi = 4D^2$. 
    Finally, we evaluate the energy $e$ and the local magnetization $m$ of the optimized states at increasingly large $\chi$ until their convergence, with $\chi_{\textrm{obs}}=10D^2$ being sufficient.

    \begin{figure}[tb]
        \centering
        \includegraphics[width=\columnwidth]{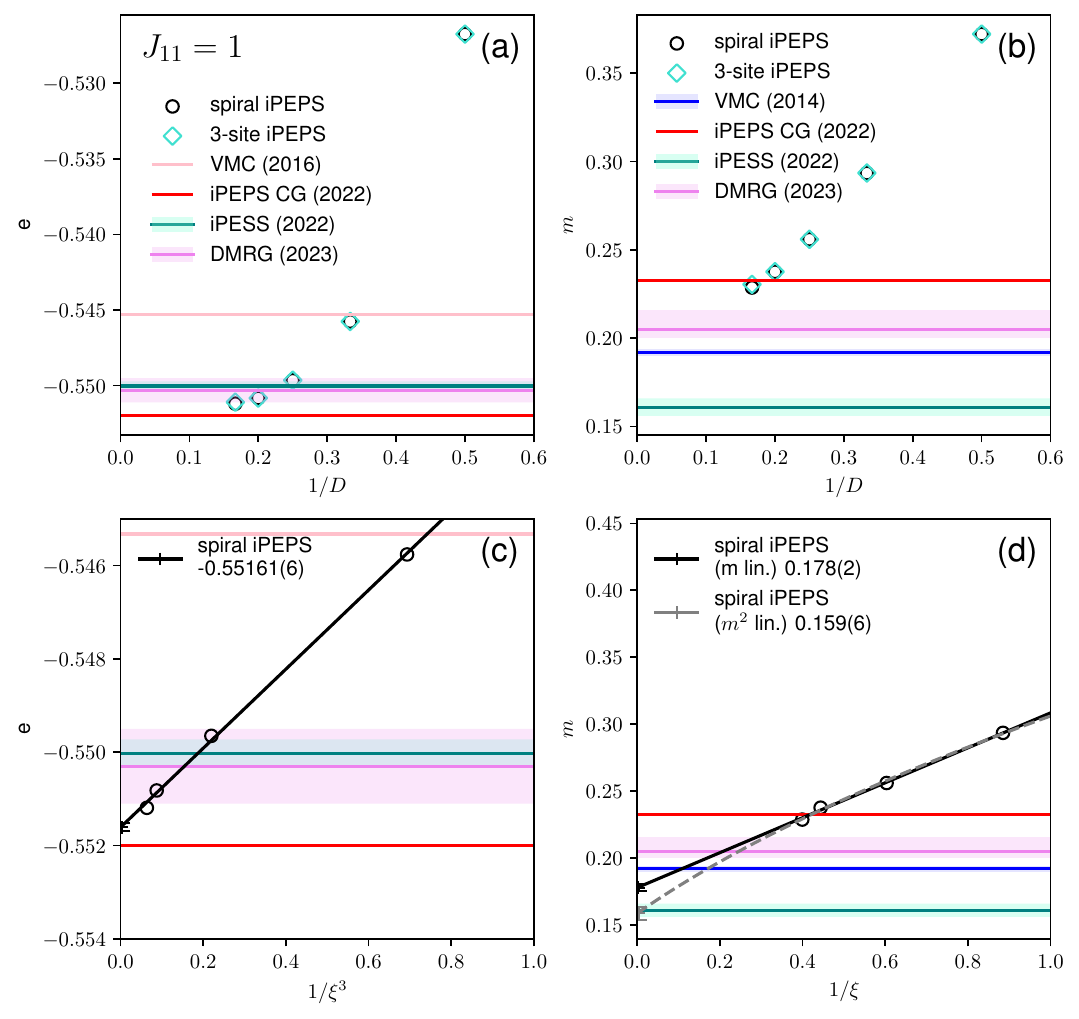}
        \caption{Isotropic triangular lattice results, $J_{11}=1$. (a)~Energy per site $e$  and (b) local magnetization $m$ of spiral iPEPS and  three-site iPEPS as a function of inverse bond dimension. (c)~Extrapolation of the spiral iPEPS energy $e$ and (d) the local magnetization $m$ with respect to the effective correlation length $\xi$ (see text). For comparison, extrapolations from DMRG~\cite{huang2023magnetization}, iPESS~\cite{li2022magnetization}, coarse-grained (CG) iPEPS~\cite{PhysRevLett.129.227201}, and VMC~\cite{PhysRevB.93.144411,doi:10.7566/JPSJ.83.093707}  are included.
        } 
            \label{fig:tlhafm} 
    \end{figure}

    First, we focus on the finite-$D$ results for the isotropic triangular case at $J_{11}=1$ in Figs.~\hyperref[fig:tlhafm]{\ref*{fig:tlhafm}(a)} and~\hyperref[fig:tlhafm]{\ref*{fig:tlhafm}(b)} and compare them with the standard iPEPS approach, where the three-sublattice 120$^\circ$ order is simulated using a unit cell with three independent tensors~\cite{niesen2018}. A spiral iPEPS at fixed $\mathbf{q}=(2\pi/3,2\pi/3)$, which is a subset of three-site iPEPS, shows an energy difference of less than $10^{-4}$ for the largest $D=6$ considered here, despite having just one third of variational parameters. 
    Similarly, the local magnetic moments of the two ansätze are also consistent. This demonstrates the ability of a spiral iPEPS to faithfully capture the 120$^\circ$ order and to do so at a computational cost of only single-site iPEPS. Moreover, already at $D=6$ the optimized iPEPS exhibits a lower energy, $e(D=6)=-0.55119$, than extrapolations from the recent state-of-the-art DMRG simulations on cylinders of width up to 12~\cite{huang2023magnetization}, with extrapolated energy $e_{\textrm{DMRG}}=-0.5503(8)$. This highlights the efficiency of the iPEPS Ansatz, using only $2\times6^4$ parameters, in stark contrast to the billions of parameters used in the largest DMRG simulations.
    
\begin{figure*}[t]
        \centering
        \includegraphics[width=\textwidth]{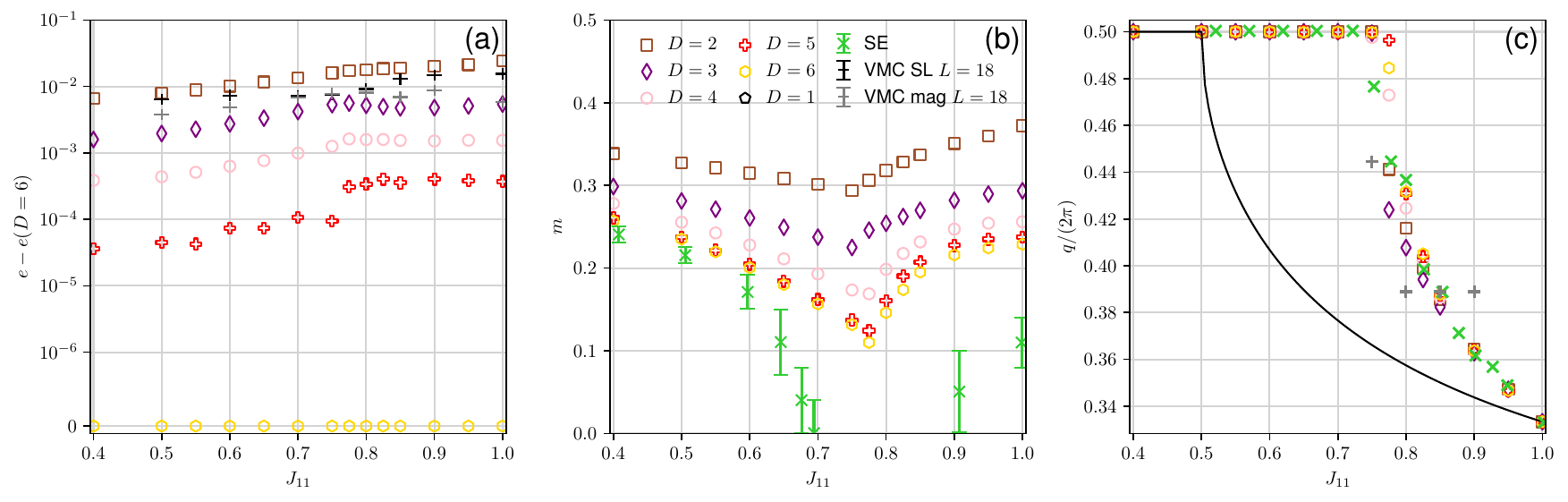}
        \caption{Results for the anisotropic triangular Heisenberg model from spiral iPEPS with bond dimensions ranging from $D=1$ (equivalent to a mean field solution) up to $D=6$. 
        (a) Energy differences with respect to $D=6$. (b) Local magnetization $m$. (c) Optimal wave vector $\mathbf{q}=(q, q)$, with $\mathbf{q}=(\pi,\pi)$ in the  N\'eel phase and $\mathbf{q}=(2\pi/3,2\pi/3)$ in the coplanar $120^\circ$ ordered phase realized at $J_{11}=1$.
        The VMC data (pluses) and SE data (crosses) are taken from Ref.~\cite{PhysRevB.93.085111} and Ref.~\cite{PhysRevB.59.14367}, respectively.
        }
        \label{fig:spiralipeps} 
    \end{figure*}

    Next, we present the results at finite anisotropy in the interval $J_{11}\in[0.4,1]$ in Fig.~\ref{fig:spiralipeps}.
    In terms of variational energy, spiral iPEPS show a 
    systematic improvement with $D$ across the full phase diagram, and exhibit better variational energies than VMC, see Fig.~\hyperref[fig:spiralipeps]{\ref*{fig:spiralipeps}(a)}. 
    The iPEPS local magnetization $m$ in Fig.~\hyperref[fig:spiralipeps]{\ref*{fig:spiralipeps}(b)} closely follows the SE data up to $J_{11}\sim0.6$, beyond which SE predicts that the N\'eel order melts rapidly as the system dimerizes into a columnar valence bond crystal (VBC)~\cite{PhysRevB.59.14367}. On the contrary, such behavior is disputed by VMC calculations~\cite{PhysRevB.93.085111}, which do not report VBC order and instead point to a QSL arising in the narrow, highly frustrated region $J_{11}\in[0.7,0.8]$. We will address this region in greater detail using a systematic scaling analysis of the iPEPS data further below. 
    
    In agreement with SE, the optimal wave vector $\mathbf{q}$, shown in  Fig.~\hyperref[fig:spiralipeps]{\ref*{fig:spiralipeps}(c)}, remains at $\mathbf{q}=(\pi,\pi)$ up to $J_{11}\sim0.75$, well beyond the $J_{11,c}^{MF}=0.5$ suggested by  the mean-field solution -- a feature which can be understood in terms of an order-by-disorder mechanism~\cite{PhysRevB.59.14367}. 
    Importantly, for $J_{11}\geq0.8$ the spiral iPEPS give a compelling picture of the spiral phase, with smoothly varying optimal wave vector $\mathbf{q}$ as $J_{11}$ is increased up to $J_{11}=1$. 
    Let us highlight that contrary to finite-size methods such as VMC or DMRG, we are not bound by the physical size of the system and thus can explore an entire range of spirals even if their periods span thousands of lattice sites, without increasing computational complexity.
    
    \emph{{Finite correlation-length scaling.}}---
    Now we turn to assess the phase diagram in the infinite-$D$ limit by a finite correlation-length scaling (FCLS) analysis~\cite{Corboz2018,Rader2018}, which is similar to  conventional finite-size scaling, but where the system size is replaced by the iPEPS correlation length $\xi$ as the effective length scale.
    Following FCLS, we extrapolate the energy $e$ and local magnetization $m$ as
    \begin{align}
    e(1/\xi)&=e_0 + \frac{a}{\xi^3} + O\left(\frac{1}{\xi^4}\right), \\
    m^p(1/\xi)&=m^p_0 + \frac{\alpha_p}{\xi} + O\left(\frac{1}{\xi^2}\right)
    \label{eqn:fcls_m}
    \end{align}
    where the case of $p=2$ is the analytical form of the leading finite-size corrections  in the N\'eel ordered phase~\cite{hasenfratz93}. For the spiral phase we examine both $p=1$ and $p=2$, since the analytical form of finite-size corrections in the spiral phase, and the 120$^\circ$ ordered phase in particular, is currently not known. 
    The iPEPS correlation functions and their correlation lengths are determined by the iPEPS transfer matrix, which we describe in detail in the SM~\cite{SM}.

    First, we present a detailed example of FCLS applied to the energy and order parameter in the isotropic case ($J_{11}=1$) in Figs.~\hyperref[fig:tlhafm]{\ref*{fig:tlhafm}(c)} and ~\hyperref[fig:tlhafm]{\ref*{fig:tlhafm}(d)}, respectively. The extrapolated energy $e_0=-0.55160(9)$ lies in between the extrapolated data from DMRG~\cite{PhysRevB.93.144411} and the coarse-grained iPEPS~\cite{PhysRevLett.129.227201}. The local magnetization  
    obtained from an extrapolation of $m$, i.e., the case of $p=1$, yields $m_0=0.178(2)$ which is roughly $10\%$ lower than previous DMRG~\cite{white2007order,huang2023magnetization} and VMC~\cite{doi:10.7566/JPSJ.83.093707} estimates. The extrapolation of $m^2$ (i.e., $p=2$)  leads to an even lower estimate, $m_0=0.159(5)$.
    
\begin{figure}[tb]
        \centering
        \includegraphics[width=\columnwidth]{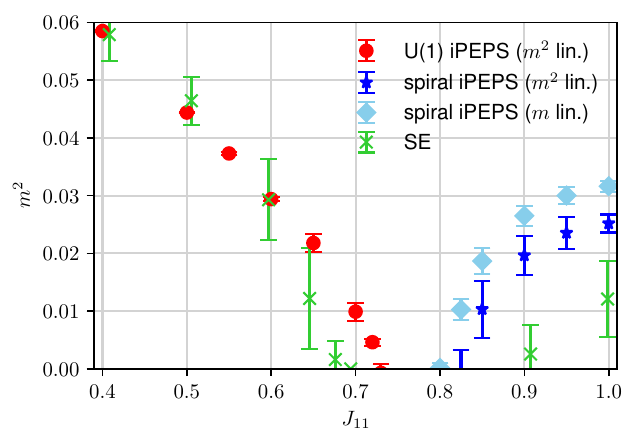}
        \caption{Extrapolated values of the squared local 
        magnetization $m^2$ for $J_{11}\leq0.75$ (filled circles) based on FCLS of a U(1)-symmetric bipartite iPEPS with fixed $\bf{q}=(\pi,\pi)$ order. For $J_{11}\geq0.8$, we use a spiral iPEPS Ansatz and plot estimates of $m^2$ based on an extrapolation of $m$  (diamonds) or its square (stars). In all cases a first-order polynomial is used to extrapolate the data. SE data (crosses) are taken from Ref.~\cite{PhysRevB.59.14367}.}
        \label{fig:phase_diag} 
    \end{figure}

    Now we proceed with the scaling analysis over the full interval $J_{11}\in[0.4,1]$.
    Building on the spiral iPEPS results, which suggest that the dominant magnetic correlations retain collinear $\mathbf{q}=(\pi,\pi)$ character below $J_{11}\sim0.8$, we utilize a more efficient bipartite iPEPS Ansatz in this region, taking advantage of  
    the remnant U(1) symmetry. This allows us to reach higher $D$ and $\chi$. Following an identical optimization protocol as for spiral iPEPS, these states were optimized up to $D=7$ at environment dimensions $\chi=4D^2$. The FCLS analysis shown in Fig.~\ref{fig:phase_diag} reveals that the N\'eel order vanishes at $J_{11,c1}\sim0.73(1)$. At larger values, $J_{11,c2}\sim0.80(2)$, the system becomes ordered again as the spin spirals form. Importantly, throughout the N\'eel phase the local magnetization data are well captured by the scaling formula of Eq.~\ref{eqn:fcls_m} with $p=2$, while in the spiral phase, both $p=1$ and $p=2$ formulas provide plausible fits to the $D\leq6$ spiral iPEPS data. Details on the extrapolations are provided in the SM~\cite{SM}. 
    
    Finally, we discuss the nature of the paramagnetic phase emerging between the magnetically  ordered phases. According to our results, it has short-range $\mathbf{q}=(\pi,\pi)$ antiferromagnetic correlations and does not break the translational nor the point group symmetries of the anisotropic triangular lattice. Besides this, using a standard iPEPS with a $2\times1$ unit cell,  we find the competing dimerized state suggested earlier by SE~\cite{PhysRevB.59.14367} to be energetically disfavored (see SM~\cite{SM}). Together, these results provide robust evidence for the presence of a QSL   between the N\'eel and spin spiral phases, as previously suggested by VMC~\cite{PhysRevB.93.085111}, which however was limited in the accessible values of $\bf{q}$.

    %
    
    \emph{{Conclusions.}} -- Our work extends the tensor network toolbox by spiral iPEPS -- an Ansatz that is capable of simulating magnetic states ranging from N\'eel order to any $\mathbf{q}$-vector spin spiral - while keeping the computational cost at the level of  single-site iPEPS. This paves the way for iPEPS simulations of incommensurate orders in two dimensions, which would typically require prohibitively large unit cells, and it enables investigating  such orders without the spurious effects associated with the  finite-size geometry. We have demonstrated the performance of a spiral iPEPS for the Heisenberg model on the anisotropic triangular lattice, faithfully capturing all phases in its phase diagram, from the N\'eel to the spin spiral phase. In particular, the FCLS analysis of the data from a spiral iPEPS and biparite U(1)-symmetric iPEPS revealed a QSL in the region $J_{11}\in[0.73(1),0.80(2)]$. At the same time, the  scaling analysis for the $J_{11}\geq0.8$ spiral phase and isotropic triangular lattice model at $J_{11}=1$ calls for further investigation of the analytical form of the finite-size corrections. We envision the applications of a spiral iPEPS beyond spin-1/2 systems to SU(N) models, where some of the symmetry-broken phases~\cite{toth2010,bauer2011-su3} could be analogously expressed through the action of an appropriate unitary $U(\mathbf{q})$.
    Finally, the combination of a spiral iPEPS with a standard multisite iPEPS offers a powerful approach to study the competition between spiral states and translational symmetry broken states such as valence bond crystals, or certain QSLs which do not allow for a single-site iPEPS description.



    \par\noindent\emph{\textbf{Acknowledgments---}} %
    We thank Hiroshi Ueda for bringing to our attention Ref.~\cite{ueda2012}. We acknowledge support from the European Research Council (ERC) under the European Union's Horizon 2020 research and innovation programme (Grant Agreement No. 101001604). All simulations were conducted with \textsc{peps-torch}~\cite{pepstorch} and \textsc{YASTN}~\cite{yastn} tensor network libraries. We thank SURF (www.surf.nl) for the support in using the National Supercomputers Lisa and Snellius.
    
    \bibliography{bibliography,refs}

\begin{thebibliography}{61}%
\makeatletter
\providecommand \@ifxundefined [1]{%
 \@ifx{#1\undefined}
}%
\providecommand \@ifnum [1]{%
 \ifnum #1\expandafter \@firstoftwo
 \else \expandafter \@secondoftwo
 \fi
}%
\providecommand \@ifx [1]{%
 \ifx #1\expandafter \@firstoftwo
 \else \expandafter \@secondoftwo
 \fi
}%
\providecommand \natexlab [1]{#1}%
\providecommand \enquote  [1]{``#1''}%
\providecommand \bibnamefont  [1]{#1}%
\providecommand \bibfnamefont [1]{#1}%
\providecommand \citenamefont [1]{#1}%
\providecommand \href@noop [0]{\@secondoftwo}%
\providecommand \href [0]{\begingroup \@sanitize@url \@href}%
\providecommand \@href[1]{\@@startlink{#1}\@@href}%
\providecommand \@@href[1]{\endgroup#1\@@endlink}%
\providecommand \@sanitize@url [0]{\catcode `\\12\catcode `\$12\catcode
  `\&12\catcode `\#12\catcode `\^12\catcode `\_12\catcode `\%12\relax}%
\providecommand \@@startlink[1]{}%
\providecommand \@@endlink[0]{}%
\providecommand \url  [0]{\begingroup\@sanitize@url \@url }%
\providecommand \@url [1]{\endgroup\@href {#1}{\urlprefix }}%
\providecommand \urlprefix  [0]{URL }%
\providecommand \Eprint [0]{\href }%
\providecommand \doibase [0]{https://doi.org/}%
\providecommand \selectlanguage [0]{\@gobble}%
\providecommand \bibinfo  [0]{\@secondoftwo}%
\providecommand \bibfield  [0]{\@secondoftwo}%
\providecommand \translation [1]{[#1]}%
\providecommand \BibitemOpen [0]{}%
\providecommand \bibitemStop [0]{}%
\providecommand \bibitemNoStop [0]{.\EOS\space}%
\providecommand \EOS [0]{\spacefactor3000\relax}%
\providecommand \BibitemShut  [1]{\csname bibitem#1\endcsname}%
\let\auto@bib@innerbib\@empty
\bibitem [{\citenamefont {Wen}\ \emph {et~al.}(1989)\citenamefont {Wen},
  \citenamefont {Wilczek},\ and\ \citenamefont {Zee}}]{Wen1989}%
  \BibitemOpen
  \bibfield  {author} {\bibinfo {author} {\bibfnamefont {X.~G.}\ \bibnamefont
  {Wen}}, \bibinfo {author} {\bibfnamefont {F.}~\bibnamefont {Wilczek}},\ and\
  \bibinfo {author} {\bibfnamefont {A.}~\bibnamefont {Zee}},\ }\bibfield
  {title} {\bibinfo {title} {Chiral spin states and superconductivity},\ }\href
  {https://doi.org/10.1103/PhysRevB.39.11413} {\bibfield  {journal} {\bibinfo
  {journal} {Phys. Rev. B}\ }\textbf {\bibinfo {volume} {39}},\ \bibinfo
  {pages} {11413} (\bibinfo {year} {1989})}\BibitemShut {NoStop}%
\bibitem [{\citenamefont {Balents}(2010)}]{balents2010}%
  \BibitemOpen
  \bibfield  {author} {\bibinfo {author} {\bibfnamefont {L.}~\bibnamefont
  {Balents}},\ }\bibfield  {title} {\bibinfo {title} {Spin liquids in
  frustrated magnets},\ }\href
  {https://doi.org/https://doi.org/10.1038/nature08917} {\bibfield  {journal}
  {\bibinfo  {journal} {Nature}\ }\textbf {\bibinfo {volume} {464}},\ \bibinfo
  {pages} {199} (\bibinfo {year} {2010})}\BibitemShut {NoStop}%
\bibitem [{\citenamefont {White}(1992)}]{white1992}%
  \BibitemOpen
  \bibfield  {author} {\bibinfo {author} {\bibfnamefont {S.~R.}\ \bibnamefont
  {White}},\ }\bibfield  {title} {\bibinfo {title} {Density matrix formulation
  for quantum renormalization groups},\ }\href
  {https://doi.org/10.1103/PhysRevLett.69.2863} {\bibfield  {journal} {\bibinfo
   {journal} {Phys. Rev. Lett.}\ }\textbf {\bibinfo {volume} {69}},\ \bibinfo
  {pages} {2863} (\bibinfo {year} {1992})}\BibitemShut {NoStop}%
\bibitem [{\citenamefont {\"Ostlund}\ and\ \citenamefont
  {Rommer}(1995)}]{ostlund1995}%
  \BibitemOpen
  \bibfield  {author} {\bibinfo {author} {\bibfnamefont {S.}~\bibnamefont
  {\"Ostlund}}\ and\ \bibinfo {author} {\bibfnamefont {S.}~\bibnamefont
  {Rommer}},\ }\bibfield  {title} {\bibinfo {title} {Thermodynamic limit of
  density matrix renormalization},\ }\href
  {https://doi.org/10.1103/PhysRevLett.75.3537} {\bibfield  {journal} {\bibinfo
   {journal} {Phys. Rev. Lett.}\ }\textbf {\bibinfo {volume} {75}},\ \bibinfo
  {pages} {3537} (\bibinfo {year} {1995})}\BibitemShut {NoStop}%
\bibitem [{\citenamefont {Perez-Garcia}\ \emph {et~al.}(2007)\citenamefont
  {Perez-Garcia}, \citenamefont {Verstraete}, \citenamefont {Wolf},\ and\
  \citenamefont {Cirac}}]{perez2007}%
  \BibitemOpen
  \bibfield  {author} {\bibinfo {author} {\bibfnamefont {D.}~\bibnamefont
  {Perez-Garcia}}, \bibinfo {author} {\bibfnamefont {F.}~\bibnamefont
  {Verstraete}}, \bibinfo {author} {\bibfnamefont {M.}~\bibnamefont {Wolf}},\
  and\ \bibinfo {author} {\bibfnamefont {J.~I.}\ \bibnamefont {Cirac}},\
  }\href@noop {} {\bibinfo {title} {Matrix product state representations}}
  (\bibinfo {year} {2007}),\ \Eprint {https://arxiv.org/abs/quant-ph/0608197}
  {arXiv:quant-ph/0608197 [quant-ph]} \BibitemShut {NoStop}%
\bibitem [{\citenamefont {Verstraete}\ and\ \citenamefont
  {Cirac}(2004)}]{Verstraete2004}%
  \BibitemOpen
  \bibfield  {author} {\bibinfo {author} {\bibfnamefont {F.}~\bibnamefont
  {Verstraete}}\ and\ \bibinfo {author} {\bibfnamefont {J.~I.}\ \bibnamefont
  {Cirac}},\ }\bibfield  {title} {\bibinfo {title} {Renormalization algorithms
  for quantum-many body systems in two and higher dimensions},\ }\href
  {https://arxiv.org/abs/cond-mat/0407066} {\bibfield  {journal} {\bibinfo
  {journal} {arXiv:cond-mat/04070663}\ } (\bibinfo {year} {2004})}\BibitemShut
  {NoStop}%
\bibitem [{\citenamefont {Nishio}\ \emph {et~al.}(2004)\citenamefont {Nishio},
  \citenamefont {Maeshima}, \citenamefont {Gendiar},\ and\ \citenamefont
  {Nishino}}]{nishio2004}%
  \BibitemOpen
  \bibfield  {author} {\bibinfo {author} {\bibfnamefont {Y.}~\bibnamefont
  {Nishio}}, \bibinfo {author} {\bibfnamefont {N.}~\bibnamefont {Maeshima}},
  \bibinfo {author} {\bibfnamefont {A.}~\bibnamefont {Gendiar}},\ and\ \bibinfo
  {author} {\bibfnamefont {T.}~\bibnamefont {Nishino}},\ }\bibfield  {title}
  {\bibinfo {title} {{Tensor Product Variational Formulation for Quantum
  Systems}},\ }\href@noop {} {\bibfield  {journal} {\bibinfo  {journal}
  {Preprint}\ } (\bibinfo {year} {2004})},\ \Eprint
  {https://arxiv.org/abs/cond-mat/0401115} {arXiv:cond-mat/0401115}
  \BibitemShut {NoStop}%
\bibitem [{\citenamefont {Jordan}\ \emph {et~al.}(2008)\citenamefont {Jordan},
  \citenamefont {Or\'us}, \citenamefont {Vidal}, \citenamefont {Verstraete},\
  and\ \citenamefont {Cirac}}]{Jordan2008}%
  \BibitemOpen
  \bibfield  {author} {\bibinfo {author} {\bibfnamefont {J.}~\bibnamefont
  {Jordan}}, \bibinfo {author} {\bibfnamefont {R.}~\bibnamefont {Or\'us}},
  \bibinfo {author} {\bibfnamefont {G.}~\bibnamefont {Vidal}}, \bibinfo
  {author} {\bibfnamefont {F.}~\bibnamefont {Verstraete}},\ and\ \bibinfo
  {author} {\bibfnamefont {J.~I.}\ \bibnamefont {Cirac}},\ }\bibfield  {title}
  {\bibinfo {title} {Classical simulation of infinite-size quantum lattice
  systems in two spatial dimensions},\ }\href
  {https://doi.org/10.1103/PhysRevLett.101.250602} {\bibfield  {journal}
  {\bibinfo  {journal} {Phys. Rev. Lett.}\ }\textbf {\bibinfo {volume} {101}},\
  \bibinfo {pages} {250602} (\bibinfo {year} {2008})}\BibitemShut {NoStop}%
\bibitem [{\citenamefont {Zheng}\ \emph {et~al.}(2017)\citenamefont {Zheng},
  \citenamefont {Chung}, \citenamefont {Corboz}, \citenamefont {Ehlers},
  \citenamefont {Qin}, \citenamefont {Noack}, \citenamefont {Shi},
  \citenamefont {White}, \citenamefont {Zhang},\ and\ \citenamefont
  {Chan}}]{zheng17}%
  \BibitemOpen
  \bibfield  {author} {\bibinfo {author} {\bibfnamefont {B.-X.}\ \bibnamefont
  {Zheng}}, \bibinfo {author} {\bibfnamefont {C.-M.}\ \bibnamefont {Chung}},
  \bibinfo {author} {\bibfnamefont {P.}~\bibnamefont {Corboz}}, \bibinfo
  {author} {\bibfnamefont {G.}~\bibnamefont {Ehlers}}, \bibinfo {author}
  {\bibfnamefont {M.-P.}\ \bibnamefont {Qin}}, \bibinfo {author} {\bibfnamefont
  {R.~M.}\ \bibnamefont {Noack}}, \bibinfo {author} {\bibfnamefont
  {H.}~\bibnamefont {Shi}}, \bibinfo {author} {\bibfnamefont {S.~R.}\
  \bibnamefont {White}}, \bibinfo {author} {\bibfnamefont {S.}~\bibnamefont
  {Zhang}},\ and\ \bibinfo {author} {\bibfnamefont {G.~K.-L.}\ \bibnamefont
  {Chan}},\ }\bibfield  {title} {{\selectlanguage {english}\bibinfo {title}
  {Stripe order in the underdoped region of the two-dimensional {Hubbard}
  model}},\ }\href {https://doi.org/10.1126/science.aam7127} {\bibfield
  {journal} {\bibinfo  {journal} {Science}\ }\textbf {\bibinfo {volume}
  {358}},\ \bibinfo {pages} {1155} (\bibinfo {year} {2017})}\BibitemShut
  {NoStop}%
\bibitem [{\citenamefont {Ponsioen}\ \emph {et~al.}(2019)\citenamefont
  {Ponsioen}, \citenamefont {Chung},\ and\ \citenamefont
  {Corboz}}]{ponsioen19}%
  \BibitemOpen
  \bibfield  {author} {\bibinfo {author} {\bibfnamefont {B.}~\bibnamefont
  {Ponsioen}}, \bibinfo {author} {\bibfnamefont {S.~S.}\ \bibnamefont
  {Chung}},\ and\ \bibinfo {author} {\bibfnamefont {P.}~\bibnamefont
  {Corboz}},\ }\bibfield  {title} {\bibinfo {title} {Period 4 stripe in the
  extended two-dimensional {Hubbard} model},\ }\href
  {https://doi.org/10.1103/PhysRevB.100.195141} {\bibfield  {journal} {\bibinfo
   {journal} {Phys. Rev. B}\ }\textbf {\bibinfo {volume} {100}},\ \bibinfo
  {pages} {195141} (\bibinfo {year} {2019})}\BibitemShut {NoStop}%
\bibitem [{\citenamefont {Zhang}\ \emph
  {et~al.}(2023{\natexlab{a}})\citenamefont {Zhang}, \citenamefont {Li},\ and\
  \citenamefont {von Delft}}]{zhang23b}%
  \BibitemOpen
  \bibfield  {author} {\bibinfo {author} {\bibfnamefont {C.}~\bibnamefont
  {Zhang}}, \bibinfo {author} {\bibfnamefont {J.-W.}\ \bibnamefont {Li}},\ and\
  \bibinfo {author} {\bibfnamefont {J.}~\bibnamefont {von Delft}},\ }\href
  {https://doi.org/10.48550/arXiv.2307.14835} {\bibinfo {title}
  {Frustration-{Induced} {Superconductivity} in the $t$-$t'$ {Hubbard}
  {Model}}} (\bibinfo {year} {2023}{\natexlab{a}}),\ \bibinfo {note}
  {arXiv:2307.14835 [cond-mat]}\BibitemShut {NoStop}%
\bibitem [{\citenamefont {Ponsioen}\ \emph {et~al.}(2023)\citenamefont
  {Ponsioen}, \citenamefont {Chung},\ and\ \citenamefont
  {Corboz}}]{ponsioen2023superconducting}%
  \BibitemOpen
  \bibfield  {author} {\bibinfo {author} {\bibfnamefont {B.}~\bibnamefont
  {Ponsioen}}, \bibinfo {author} {\bibfnamefont {S.~S.}\ \bibnamefont
  {Chung}},\ and\ \bibinfo {author} {\bibfnamefont {P.}~\bibnamefont
  {Corboz}},\ }\bibfield  {title} {\bibinfo {title} {Superconducting stripes in
  the hole-doped three-band hubbard model},\ }\href
  {https://doi.org/10.1103/PhysRevB.108.205154} {\bibfield  {journal} {\bibinfo
   {journal} {Phys. Rev. B}\ }\textbf {\bibinfo {volume} {108}},\ \bibinfo
  {pages} {205154} (\bibinfo {year} {2023})}\BibitemShut {NoStop}%
\bibitem [{\citenamefont {Niesen}\ and\ \citenamefont
  {Corboz}(2017)}]{niesen17}%
  \BibitemOpen
  \bibfield  {author} {\bibinfo {author} {\bibfnamefont {I.}~\bibnamefont
  {Niesen}}\ and\ \bibinfo {author} {\bibfnamefont {P.}~\bibnamefont
  {Corboz}},\ }\bibfield  {title} {\bibinfo {title} {Emergent {Haldane} phase
  in the \${S}=1\$ bilinear-biquadratic {Heisenberg} model on the square
  lattice},\ }\href {https://doi.org/10.1103/PhysRevB.95.180404} {\bibfield
  {journal} {\bibinfo  {journal} {Phys. Rev. B}\ }\textbf {\bibinfo {volume}
  {95}},\ \bibinfo {pages} {180404(R)} (\bibinfo {year} {2017})}\BibitemShut
  {NoStop}%
\bibitem [{\citenamefont {Chen}\ \emph {et~al.}(2018)\citenamefont {Chen},
  \citenamefont {Vanderstraeten}, \citenamefont {Capponi},\ and\ \citenamefont
  {Poilblanc}}]{chen18}%
  \BibitemOpen
  \bibfield  {author} {\bibinfo {author} {\bibfnamefont {J.-Y.}\ \bibnamefont
  {Chen}}, \bibinfo {author} {\bibfnamefont {L.}~\bibnamefont
  {Vanderstraeten}}, \bibinfo {author} {\bibfnamefont {S.}~\bibnamefont
  {Capponi}},\ and\ \bibinfo {author} {\bibfnamefont {D.}~\bibnamefont
  {Poilblanc}},\ }\bibfield  {title} {\bibinfo {title} {Non-{Abelian} chiral
  spin liquid in a quantum antiferromagnet revealed by an {iPEPS} study},\
  }\href {https://doi.org/10.1103/PhysRevB.98.184409} {\bibfield  {journal}
  {\bibinfo  {journal} {Phys. Rev. B}\ }\textbf {\bibinfo {volume} {98}},\
  \bibinfo {pages} {184409} (\bibinfo {year} {2018})}\BibitemShut {NoStop}%
\bibitem [{\citenamefont {Hasik}\ \emph
  {et~al.}(2021{\natexlab{a}})\citenamefont {Hasik}, \citenamefont
  {Poilblanc},\ and\ \citenamefont {Becca}}]{hasik21}%
  \BibitemOpen
  \bibfield  {author} {\bibinfo {author} {\bibfnamefont {J.}~\bibnamefont
  {Hasik}}, \bibinfo {author} {\bibfnamefont {D.}~\bibnamefont {Poilblanc}},\
  and\ \bibinfo {author} {\bibfnamefont {F.}~\bibnamefont {Becca}},\ }\bibfield
   {title} {{\selectlanguage {english}\bibinfo {title} {Investigation of the
  {N{\'e}el} phase of the frustrated {Heisenberg} antiferromagnet by
  differentiable symmetric tensor networks}},\ }\href
  {https://doi.org/10.21468/SciPostPhys.10.1.012} {\bibfield  {journal}
  {\bibinfo  {journal} {SciPost Physics}\ }\textbf {\bibinfo {volume} {10}},\
  \bibinfo {pages} {012} (\bibinfo {year} {2021}{\natexlab{a}})}\BibitemShut
  {NoStop}%
\bibitem [{\citenamefont {Liu}\ \emph {et~al.}(2022)\citenamefont {Liu},
  \citenamefont {Hasik}, \citenamefont {Gong}, \citenamefont {Poilblanc},
  \citenamefont {Chen},\ and\ \citenamefont {Gu}}]{liu22b}%
  \BibitemOpen
  \bibfield  {author} {\bibinfo {author} {\bibfnamefont {W.-Y.}\ \bibnamefont
  {Liu}}, \bibinfo {author} {\bibfnamefont {J.}~\bibnamefont {Hasik}}, \bibinfo
  {author} {\bibfnamefont {S.-S.}\ \bibnamefont {Gong}}, \bibinfo {author}
  {\bibfnamefont {D.}~\bibnamefont {Poilblanc}}, \bibinfo {author}
  {\bibfnamefont {W.-Q.}\ \bibnamefont {Chen}},\ and\ \bibinfo {author}
  {\bibfnamefont {Z.-C.}\ \bibnamefont {Gu}},\ }\bibfield  {title} {\bibinfo
  {title} {Emergence of {Gapless} {Quantum} {Spin} {Liquid} from {Deconfined}
  {Quantum} {Critical} {Point}},\ }\href
  {https://doi.org/10.1103/PhysRevX.12.031039} {\bibfield  {journal} {\bibinfo
  {journal} {Phys. Rev. X}\ }\textbf {\bibinfo {volume} {12}},\ \bibinfo
  {pages} {031039} (\bibinfo {year} {2022})}\BibitemShut {NoStop}%
\bibitem [{\citenamefont {Hasik}\ \emph {et~al.}(2022)\citenamefont {Hasik},
  \citenamefont {Van~Damme}, \citenamefont {Poilblanc},\ and\ \citenamefont
  {Vanderstraeten}}]{hasik22}%
  \BibitemOpen
  \bibfield  {author} {\bibinfo {author} {\bibfnamefont {J.}~\bibnamefont
  {Hasik}}, \bibinfo {author} {\bibfnamefont {M.}~\bibnamefont {Van~Damme}},
  \bibinfo {author} {\bibfnamefont {D.}~\bibnamefont {Poilblanc}},\ and\
  \bibinfo {author} {\bibfnamefont {L.}~\bibnamefont {Vanderstraeten}},\
  }\bibfield  {title} {\bibinfo {title} {Simulating {Chiral} {Spin} {Liquids}
  with {Projected} {Entangled}-{Pair} {States}},\ }\href
  {https://doi.org/10.1103/PhysRevLett.129.177201} {\bibfield  {journal}
  {\bibinfo  {journal} {Phys. Rev. Lett.}\ }\textbf {\bibinfo {volume} {129}},\
  \bibinfo {pages} {177201} (\bibinfo {year} {2022})}\BibitemShut {NoStop}%
\bibitem [{\citenamefont {Liao}\ \emph {et~al.}(2017)\citenamefont {Liao},
  \citenamefont {Xie}, \citenamefont {Chen}, \citenamefont {Liu}, \citenamefont
  {Xie}, \citenamefont {Huang}, \citenamefont {Normand},\ and\ \citenamefont
  {Xiang}}]{PhysRevLett.118.137202}%
  \BibitemOpen
  \bibfield  {author} {\bibinfo {author} {\bibfnamefont {H.~J.}\ \bibnamefont
  {Liao}}, \bibinfo {author} {\bibfnamefont {Z.~Y.}\ \bibnamefont {Xie}},
  \bibinfo {author} {\bibfnamefont {J.}~\bibnamefont {Chen}}, \bibinfo {author}
  {\bibfnamefont {Z.~Y.}\ \bibnamefont {Liu}}, \bibinfo {author} {\bibfnamefont
  {H.~D.}\ \bibnamefont {Xie}}, \bibinfo {author} {\bibfnamefont {R.~Z.}\
  \bibnamefont {Huang}}, \bibinfo {author} {\bibfnamefont {B.}~\bibnamefont
  {Normand}},\ and\ \bibinfo {author} {\bibfnamefont {T.}~\bibnamefont
  {Xiang}},\ }\bibfield  {title} {\bibinfo {title} {Gapless spin-liquid ground
  state in the $s=1/2$ kagome antiferromagnet},\ }\href
  {https://doi.org/10.1103/PhysRevLett.118.137202} {\bibfield  {journal}
  {\bibinfo  {journal} {Phys. Rev. Lett.}\ }\textbf {\bibinfo {volume} {118}},\
  \bibinfo {pages} {137202} (\bibinfo {year} {2017})}\BibitemShut {NoStop}%
\bibitem [{\citenamefont {Kshetrimayum}\ \emph {et~al.}(2020)\citenamefont
  {Kshetrimayum}, \citenamefont {Balz}, \citenamefont {Lake},\ and\
  \citenamefont {Eisert}}]{kshetrimayum19b}%
  \BibitemOpen
  \bibfield  {author} {\bibinfo {author} {\bibfnamefont {A.}~\bibnamefont
  {Kshetrimayum}}, \bibinfo {author} {\bibfnamefont {C.}~\bibnamefont {Balz}},
  \bibinfo {author} {\bibfnamefont {B.}~\bibnamefont {Lake}},\ and\ \bibinfo
  {author} {\bibfnamefont {J.}~\bibnamefont {Eisert}},\ }\bibfield  {title}
  {\bibinfo {title} {Tensor network investigation of the double layer kagome
  compound $ca_{10}cr_{7}o_{28}$},\ }\href
  {https://doi.org/https://doi.org/10.1016/j.aop.2020.168292} {\bibfield
  {journal} {\bibinfo  {journal} {Annals of Physics}\ }\textbf {\bibinfo
  {volume} {421}},\ \bibinfo {pages} {168292} (\bibinfo {year}
  {2020})}\BibitemShut {NoStop}%
\bibitem [{\citenamefont {Corboz}\ and\ \citenamefont
  {Mila}(2013)}]{PhysRevB.87.115144}%
  \BibitemOpen
  \bibfield  {author} {\bibinfo {author} {\bibfnamefont {P.}~\bibnamefont
  {Corboz}}\ and\ \bibinfo {author} {\bibfnamefont {F.}~\bibnamefont {Mila}},\
  }\bibfield  {title} {\bibinfo {title} {Tensor network study of the
  shastry-sutherland model in zero magnetic field},\ }\href
  {https://doi.org/10.1103/PhysRevB.87.115144} {\bibfield  {journal} {\bibinfo
  {journal} {Phys. Rev. B}\ }\textbf {\bibinfo {volume} {87}},\ \bibinfo
  {pages} {115144} (\bibinfo {year} {2013})}\BibitemShut {NoStop}%
\bibitem [{\citenamefont {Corboz}\ and\ \citenamefont
  {Mila}(2014)}]{corboz14_shastry}%
  \BibitemOpen
  \bibfield  {author} {\bibinfo {author} {\bibfnamefont {P.}~\bibnamefont
  {Corboz}}\ and\ \bibinfo {author} {\bibfnamefont {F.}~\bibnamefont {Mila}},\
  }\bibfield  {title} {\bibinfo {title} {Crystals of bound states in the
  magnetization plateaus of the shastry-sutherland model},\ }\href
  {https://doi.org/10.1103/PhysRevLett.112.147203} {\bibfield  {journal}
  {\bibinfo  {journal} {Phys. Rev. Lett.}\ }\textbf {\bibinfo {volume} {112}},\
  \bibinfo {pages} {147203} (\bibinfo {year} {2014})}\BibitemShut {NoStop}%
\bibitem [{\citenamefont {Lee}\ and\ \citenamefont {Kawashima}(2018)}]{lee18}%
  \BibitemOpen
  \bibfield  {author} {\bibinfo {author} {\bibfnamefont {H.-Y.}\ \bibnamefont
  {Lee}}\ and\ \bibinfo {author} {\bibfnamefont {N.}~\bibnamefont
  {Kawashima}},\ }\bibfield  {title} {\bibinfo {title} {Spin-one
  bilinear-biquadratic model on a star lattice},\ }\href
  {https://doi.org/10.1103/PhysRevB.97.205123} {\bibfield  {journal} {\bibinfo
  {journal} {Phys. Rev. B}\ }\textbf {\bibinfo {volume} {97}},\ \bibinfo
  {pages} {205123} (\bibinfo {year} {2018})}\BibitemShut {NoStop}%
\bibitem [{\citenamefont {Jahromi}\ and\ \citenamefont
  {Or{\'u}s}(2018)}]{jahromi18}%
  \BibitemOpen
  \bibfield  {author} {\bibinfo {author} {\bibfnamefont {S.~S.}\ \bibnamefont
  {Jahromi}}\ and\ \bibinfo {author} {\bibfnamefont {R.}~\bibnamefont
  {Or{\'u}s}},\ }\bibfield  {title} {\bibinfo {title} {Spin-$\frac{1}{2}$
  {Heisenberg} antiferromagnet on the star lattice: {Competing}
  valence-bond-solid phases studied by means of tensor networks},\ }\href
  {https://doi.org/10.1103/PhysRevB.98.155108} {\bibfield  {journal} {\bibinfo
  {journal} {Phys. Rev. B}\ }\textbf {\bibinfo {volume} {98}},\ \bibinfo
  {pages} {155108} (\bibinfo {year} {2018})}\BibitemShut {NoStop}%
\bibitem [{\citenamefont {Lee}\ \emph {et~al.}(2020)\citenamefont {Lee},
  \citenamefont {Kaneko}, \citenamefont {Chern}, \citenamefont {Okubo},
  \citenamefont {Yamaji}, \citenamefont {Kawashima},\ and\ \citenamefont
  {Kim}}]{lee20}%
  \BibitemOpen
  \bibfield  {author} {\bibinfo {author} {\bibfnamefont {H.-Y.}\ \bibnamefont
  {Lee}}, \bibinfo {author} {\bibfnamefont {R.}~\bibnamefont {Kaneko}},
  \bibinfo {author} {\bibfnamefont {L.~E.}\ \bibnamefont {Chern}}, \bibinfo
  {author} {\bibfnamefont {T.}~\bibnamefont {Okubo}}, \bibinfo {author}
  {\bibfnamefont {Y.}~\bibnamefont {Yamaji}}, \bibinfo {author} {\bibfnamefont
  {N.}~\bibnamefont {Kawashima}},\ and\ \bibinfo {author} {\bibfnamefont
  {Y.~B.}\ \bibnamefont {Kim}},\ }\bibfield  {title} {{\selectlanguage
  {english}\bibinfo {title} {Magnetic field induced quantum phases in a tensor
  network study of {Kitaev} magnets}},\ }\href
  {https://doi.org/10.1038/s41467-020-15320-x} {\bibfield  {journal} {\bibinfo
  {journal} {Nature Communications}\ }\textbf {\bibinfo {volume} {11}},\
  \bibinfo {pages} {1639} (\bibinfo {year} {2020})}\BibitemShut {NoStop}%
\bibitem [{\citenamefont {Zhang}\ \emph
  {et~al.}(2023{\natexlab{b}})\citenamefont {Zhang}, \citenamefont {Liang},
  \citenamefont {Liao}, \citenamefont {Li},\ and\ \citenamefont
  {Wang}}]{zhang23}%
  \BibitemOpen
  \bibfield  {author} {\bibinfo {author} {\bibfnamefont {X.-Y.}\ \bibnamefont
  {Zhang}}, \bibinfo {author} {\bibfnamefont {S.}~\bibnamefont {Liang}},
  \bibinfo {author} {\bibfnamefont {H.-J.}\ \bibnamefont {Liao}}, \bibinfo
  {author} {\bibfnamefont {W.}~\bibnamefont {Li}},\ and\ \bibinfo {author}
  {\bibfnamefont {L.}~\bibnamefont {Wang}},\ }\bibfield  {title} {\bibinfo
  {title} {Differentiable programming tensor networks for {Kitaev} magnets},\
  }\bibfield  {journal} {\bibinfo  {journal} {arXiv:2304.01551}\ }\href
  {https://doi.org/10.48550/arXiv.2304.01551} {10.48550/arXiv.2304.01551}
  (\bibinfo {year} {2023}{\natexlab{b}})\BibitemShut {NoStop}%
\bibitem [{\citenamefont {Weihong}\ \emph {et~al.}(1999)\citenamefont
  {Weihong}, \citenamefont {McKenzie},\ and\ \citenamefont
  {Singh}}]{PhysRevB.59.14367}%
  \BibitemOpen
  \bibfield  {author} {\bibinfo {author} {\bibfnamefont {Z.}~\bibnamefont
  {Weihong}}, \bibinfo {author} {\bibfnamefont {R.~H.}\ \bibnamefont
  {McKenzie}},\ and\ \bibinfo {author} {\bibfnamefont {R.~R.~P.}\ \bibnamefont
  {Singh}},\ }\bibfield  {title} {\bibinfo {title} {Phase diagram for a class
  of spin-$\frac{1}{2}$ heisenberg models interpolating between the
  square-lattice, the triangular-lattice, and the linear-chain limits},\ }\href
  {https://doi.org/10.1103/PhysRevB.59.14367} {\bibfield  {journal} {\bibinfo
  {journal} {Phys. Rev. B}\ }\textbf {\bibinfo {volume} {59}},\ \bibinfo
  {pages} {14367} (\bibinfo {year} {1999})}\BibitemShut {NoStop}%
\bibitem [{\citenamefont {Powell}\ and\ \citenamefont
  {McKenzie}(2007)}]{gutzwiller2007powell}%
  \BibitemOpen
  \bibfield  {author} {\bibinfo {author} {\bibfnamefont {B.~J.}\ \bibnamefont
  {Powell}}\ and\ \bibinfo {author} {\bibfnamefont {R.~H.}\ \bibnamefont
  {McKenzie}},\ }\bibfield  {title} {\bibinfo {title} {Symmetry of the
  superconducting order parameter in frustrated systems determined by the
  spatial anisotropy of spin correlations},\ }\href
  {https://doi.org/10.1103/PhysRevLett.98.027005} {\bibfield  {journal}
  {\bibinfo  {journal} {Phys. Rev. Lett.}\ }\textbf {\bibinfo {volume} {98}},\
  \bibinfo {pages} {027005} (\bibinfo {year} {2007})}\BibitemShut {NoStop}%
\bibitem [{\citenamefont {Bishop}\ \emph {et~al.}(2009)\citenamefont {Bishop},
  \citenamefont {Li}, \citenamefont {Farnell},\ and\ \citenamefont
  {Campbell}}]{ccluster2009bishop}%
  \BibitemOpen
  \bibfield  {author} {\bibinfo {author} {\bibfnamefont {R.~F.}\ \bibnamefont
  {Bishop}}, \bibinfo {author} {\bibfnamefont {P.~H.~Y.}\ \bibnamefont {Li}},
  \bibinfo {author} {\bibfnamefont {D.~J.~J.}\ \bibnamefont {Farnell}},\ and\
  \bibinfo {author} {\bibfnamefont {C.~E.}\ \bibnamefont {Campbell}},\
  }\bibfield  {title} {\bibinfo {title} {Magnetic order in a spin-$\frac{1}{2}$
  interpolating square-triangle heisenberg antiferromagnet},\ }\href
  {https://doi.org/10.1103/PhysRevB.79.174405} {\bibfield  {journal} {\bibinfo
  {journal} {Phys. Rev. B}\ }\textbf {\bibinfo {volume} {79}},\ \bibinfo
  {pages} {174405} (\bibinfo {year} {2009})}\BibitemShut {NoStop}%
\bibitem [{\citenamefont {Hauke}\ \emph {et~al.}(2011)\citenamefont {Hauke},
  \citenamefont {Roscilde}, \citenamefont {Murg}, \citenamefont {Cirac},\ and\
  \citenamefont {Schmied}}]{spinwave2011hauke}%
  \BibitemOpen
  \bibfield  {author} {\bibinfo {author} {\bibfnamefont {P.}~\bibnamefont
  {Hauke}}, \bibinfo {author} {\bibfnamefont {T.}~\bibnamefont {Roscilde}},
  \bibinfo {author} {\bibfnamefont {V.}~\bibnamefont {Murg}}, \bibinfo {author}
  {\bibfnamefont {J.~I.}\ \bibnamefont {Cirac}},\ and\ \bibinfo {author}
  {\bibfnamefont {R.}~\bibnamefont {Schmied}},\ }\bibfield  {title} {\bibinfo
  {title} {Modified spin-wave theory with ordering vector optimization:
  spatially anisotropic triangular lattice and $j_1$-$j_2$-$j_3$ model with
  heisenberg interactions},\ }\href
  {https://doi.org/10.1088/1367-2630/13/7/075017} {\bibfield  {journal}
  {\bibinfo  {journal} {New Journal of Physics}\ }\textbf {\bibinfo {volume}
  {13}},\ \bibinfo {pages} {075017} (\bibinfo {year} {2011})}\BibitemShut
  {NoStop}%
\bibitem [{\citenamefont {Gonzalez}\ \emph {et~al.}(2020)\citenamefont
  {Gonzalez}, \citenamefont {Ghioldi}, \citenamefont {Gazza}, \citenamefont
  {Manuel},\ and\ \citenamefont {Trumper}}]{schwinger2020gonzalez}%
  \BibitemOpen
  \bibfield  {author} {\bibinfo {author} {\bibfnamefont {M.~G.}\ \bibnamefont
  {Gonzalez}}, \bibinfo {author} {\bibfnamefont {E.~A.}\ \bibnamefont
  {Ghioldi}}, \bibinfo {author} {\bibfnamefont {C.~J.}\ \bibnamefont {Gazza}},
  \bibinfo {author} {\bibfnamefont {L.~O.}\ \bibnamefont {Manuel}},\ and\
  \bibinfo {author} {\bibfnamefont {A.~E.}\ \bibnamefont {Trumper}},\
  }\bibfield  {title} {\bibinfo {title} {Interplay between spatial anisotropy
  and next-nearest-neighbor exchange interactions in the triangular heisenberg
  model},\ }\href {https://doi.org/10.1103/PhysRevB.102.224410} {\bibfield
  {journal} {\bibinfo  {journal} {Phys. Rev. B}\ }\textbf {\bibinfo {volume}
  {102}},\ \bibinfo {pages} {224410} (\bibinfo {year} {2020})}\BibitemShut
  {NoStop}%
\bibitem [{\citenamefont {Reuther}\ and\ \citenamefont
  {Thomale}()}]{reuther11}%
  \BibitemOpen
  \bibfield  {author} {\bibinfo {author} {\bibfnamefont {J.}~\bibnamefont
  {Reuther}}\ and\ \bibinfo {author} {\bibfnamefont {R.}~\bibnamefont
  {Thomale}},\ }\bibfield  {title} {\bibinfo {title} {Functional
  renormalization group for the anisotropic triangular antiferromagnet},\
  }\href {https://doi.org/10.1103/PhysRevB.83.024402} {\bibfield  {journal}
  {\bibinfo  {journal} {Phys. Rev. B}\ }\textbf {\bibinfo {volume} {83}},\
  \bibinfo {pages} {024402}}\BibitemShut {NoStop}%
\bibitem [{\citenamefont {Weichselbaum}\ and\ \citenamefont
  {White}(2011)}]{dmrg2011weichselbaum}%
  \BibitemOpen
  \bibfield  {author} {\bibinfo {author} {\bibfnamefont {A.}~\bibnamefont
  {Weichselbaum}}\ and\ \bibinfo {author} {\bibfnamefont {S.~R.}\ \bibnamefont
  {White}},\ }\bibfield  {title} {\bibinfo {title} {Incommensurate correlations
  in the anisotropic triangular heisenberg lattice},\ }\href
  {https://doi.org/10.1103/PhysRevB.84.245130} {\bibfield  {journal} {\bibinfo
  {journal} {Phys. Rev. B}\ }\textbf {\bibinfo {volume} {84}},\ \bibinfo
  {pages} {245130} (\bibinfo {year} {2011})}\BibitemShut {NoStop}%
\bibitem [{\citenamefont {Ghorbani}\ \emph {et~al.}(2016)\citenamefont
  {Ghorbani}, \citenamefont {Tocchio},\ and\ \citenamefont
  {Becca}}]{PhysRevB.93.085111}%
  \BibitemOpen
  \bibfield  {author} {\bibinfo {author} {\bibfnamefont {E.}~\bibnamefont
  {Ghorbani}}, \bibinfo {author} {\bibfnamefont {L.~F.}\ \bibnamefont
  {Tocchio}},\ and\ \bibinfo {author} {\bibfnamefont {F.}~\bibnamefont
  {Becca}},\ }\bibfield  {title} {\bibinfo {title} {Variational wave functions
  for the $s=\frac{1}{2}$ heisenberg model on the anisotropic triangular
  lattice: Spin liquids and spiral orders},\ }\href
  {https://doi.org/10.1103/PhysRevB.93.085111} {\bibfield  {journal} {\bibinfo
  {journal} {Phys. Rev. B}\ }\textbf {\bibinfo {volume} {93}},\ \bibinfo
  {pages} {085111} (\bibinfo {year} {2016})}\BibitemShut {NoStop}%
\bibitem [{\citenamefont {Ueda}\ and\ \citenamefont
  {Maruyama}(2012)}]{ueda2012}%
  \BibitemOpen
  \bibfield  {author} {\bibinfo {author} {\bibfnamefont {H.}~\bibnamefont
  {Ueda}}\ and\ \bibinfo {author} {\bibfnamefont {I.}~\bibnamefont
  {Maruyama}},\ }\bibfield  {title} {\bibinfo {title} {Incommensurate matrix
  product state for quantum spin systems},\ }\href
  {https://doi.org/10.1103/PhysRevB.86.064438} {\bibfield  {journal} {\bibinfo
  {journal} {Phys. Rev. B}\ }\textbf {\bibinfo {volume} {86}},\ \bibinfo
  {pages} {064438} (\bibinfo {year} {2012})}\BibitemShut {NoStop}%
\bibitem [{\citenamefont {Powell}\ and\ \citenamefont
  {McKenzie}(2011)}]{powell11}%
  \BibitemOpen
  \bibfield  {author} {\bibinfo {author} {\bibfnamefont {B.~J.}\ \bibnamefont
  {Powell}}\ and\ \bibinfo {author} {\bibfnamefont {R.~H.}\ \bibnamefont
  {McKenzie}},\ }\bibfield  {title} {{\selectlanguage {english}\bibinfo {title}
  {Quantum frustration in organic {Mott} insulators: from spin liquids to
  unconventional superconductors}},\ }\href
  {https://doi.org/10.1088/0034-4885/74/5/056501} {\bibfield  {journal}
  {\bibinfo  {journal} {Rep. Prog. Phys.}\ }\textbf {\bibinfo {volume} {74}},\
  \bibinfo {pages} {056501} (\bibinfo {year} {2011})}\BibitemShut {NoStop}%
\bibitem [{\citenamefont {Kanoda}\ and\ \citenamefont {Kato}(2011)}]{kanoda11}%
  \BibitemOpen
  \bibfield  {author} {\bibinfo {author} {\bibfnamefont {K.}~\bibnamefont
  {Kanoda}}\ and\ \bibinfo {author} {\bibfnamefont {R.}~\bibnamefont {Kato}},\
  }\bibfield  {title} {{\selectlanguage {english}\bibinfo {title} {Mott
  {Physics} in {Organic} {Conductors} with {Triangular} {Lattices}}},\ }\href
  {https://doi.org/10.1146/annurev-conmatphys-062910-140521} {\bibfield
  {journal} {\bibinfo  {journal} {Annu. Rev. Condens. Matter Phys.}\ }\textbf
  {\bibinfo {volume} {2}},\ \bibinfo {pages} {167} (\bibinfo {year}
  {2011})}\BibitemShut {NoStop}%
\bibitem [{\citenamefont {Shimizu}\ \emph {et~al.}(2003)\citenamefont
  {Shimizu}, \citenamefont {Miyagawa}, \citenamefont {Kanoda}, \citenamefont
  {Maesato},\ and\ \citenamefont {Saito}}]{shimizu03}%
  \BibitemOpen
  \bibfield  {author} {\bibinfo {author} {\bibfnamefont {Y.}~\bibnamefont
  {Shimizu}}, \bibinfo {author} {\bibfnamefont {K.}~\bibnamefont {Miyagawa}},
  \bibinfo {author} {\bibfnamefont {K.}~\bibnamefont {Kanoda}}, \bibinfo
  {author} {\bibfnamefont {M.}~\bibnamefont {Maesato}},\ and\ \bibinfo {author}
  {\bibfnamefont {G.}~\bibnamefont {Saito}},\ }\bibfield  {title} {\bibinfo
  {title} {Spin {Liquid} {State} in an {Organic} {Mott} {Insulator} with a
  {Triangular} {Lattice}},\ }\href
  {https://doi.org/10.1103/PhysRevLett.91.107001} {\bibfield  {journal}
  {\bibinfo  {journal} {Phys. Rev. Lett.}\ }\textbf {\bibinfo {volume} {91}},\
  \bibinfo {pages} {107001} (\bibinfo {year} {2003})}\BibitemShut {NoStop}%
\bibitem [{\citenamefont {Manna}\ \emph {et~al.}(2010)\citenamefont {Manna},
  \citenamefont {de~Souza}, \citenamefont {Br{\"u}hl}, \citenamefont
  {Schlueter},\ and\ \citenamefont {Lang}}]{manna10}%
  \BibitemOpen
  \bibfield  {author} {\bibinfo {author} {\bibfnamefont {R.~S.}\ \bibnamefont
  {Manna}}, \bibinfo {author} {\bibfnamefont {M.}~\bibnamefont {de~Souza}},
  \bibinfo {author} {\bibfnamefont {A.}~\bibnamefont {Br{\"u}hl}}, \bibinfo
  {author} {\bibfnamefont {J.~A.}\ \bibnamefont {Schlueter}},\ and\ \bibinfo
  {author} {\bibfnamefont {M.}~\bibnamefont {Lang}},\ }\bibfield  {title}
  {\bibinfo {title} {Lattice {Effects} and {Entropy} {Release} at the
  {Low}-{Temperature} {Phase} {Transition} in the {Spin}-{Liquid} {Candidate}
  $\ensuremath{\kappa}\mathrm{\text{\ensuremath{-}}}(\mathrm{BEDT}\mathrm{\text{\ensuremath{-}}}\mathrm{TTF}{)}_{2}{\mathrm{cu}}_{2}(\mathrm{CN}{)}_{3}$},\
  }\href {https://doi.org/10.1103/PhysRevLett.104.016403} {\bibfield  {journal}
  {\bibinfo  {journal} {Phys. Rev. Lett.}\ }\textbf {\bibinfo {volume} {104}},\
  \bibinfo {pages} {016403} (\bibinfo {year} {2010})},\ \bibinfo {note}
  {publisher: American Physical Society}\BibitemShut {NoStop}%
\bibitem [{\citenamefont {Coldea}\ \emph {et~al.}(2003)\citenamefont {Coldea},
  \citenamefont {Tennant},\ and\ \citenamefont {Tylczynski}}]{coldea03}%
  \BibitemOpen
  \bibfield  {author} {\bibinfo {author} {\bibfnamefont {R.}~\bibnamefont
  {Coldea}}, \bibinfo {author} {\bibfnamefont {D.~A.}\ \bibnamefont
  {Tennant}},\ and\ \bibinfo {author} {\bibfnamefont {Z.}~\bibnamefont
  {Tylczynski}},\ }\bibfield  {title} {{\selectlanguage {english}\bibinfo
  {title} {Extended scattering continua characteristic of spin
  fractionalization in the two-dimensional frustrated quantum magnet
  {Cs}$_2${CuCl}$_4$ observed by neutron scattering}},\ }\href
  {https://doi.org/10.1103/PhysRevB.68.134424} {\bibfield  {journal} {\bibinfo
  {journal} {Phys. Rev. B}\ }\textbf {\bibinfo {volume} {68}},\ \bibinfo
  {pages} {134424} (\bibinfo {year} {2003})}\BibitemShut {NoStop}%
\bibitem [{\citenamefont {Capriotti}\ \emph {et~al.}(1999)\citenamefont
  {Capriotti}, \citenamefont {Trumper},\ and\ \citenamefont
  {Sorella}}]{PhysRevLett.82.3899}%
  \BibitemOpen
  \bibfield  {author} {\bibinfo {author} {\bibfnamefont {L.}~\bibnamefont
  {Capriotti}}, \bibinfo {author} {\bibfnamefont {A.~E.}\ \bibnamefont
  {Trumper}},\ and\ \bibinfo {author} {\bibfnamefont {S.}~\bibnamefont
  {Sorella}},\ }\bibfield  {title} {\bibinfo {title} {Long-range n\'eel order
  in the triangular heisenberg model},\ }\href
  {https://doi.org/10.1103/PhysRevLett.82.3899} {\bibfield  {journal} {\bibinfo
   {journal} {Phys. Rev. Lett.}\ }\textbf {\bibinfo {volume} {82}},\ \bibinfo
  {pages} {3899} (\bibinfo {year} {1999})}\BibitemShut {NoStop}%
\bibitem [{\citenamefont {White}\ and\ \citenamefont
  {Chernyshev}(2007)}]{white2007order}%
  \BibitemOpen
  \bibfield  {author} {\bibinfo {author} {\bibfnamefont {S.~R.}\ \bibnamefont
  {White}}\ and\ \bibinfo {author} {\bibfnamefont {A.~L.}\ \bibnamefont
  {Chernyshev}},\ }\bibfield  {title} {\bibinfo {title} {Ne\'el order in square
  and triangular lattice heisenberg models},\ }\href
  {https://doi.org/10.1103/PhysRevLett.99.127004} {\bibfield  {journal}
  {\bibinfo  {journal} {Phys. Rev. Lett.}\ }\textbf {\bibinfo {volume} {99}},\
  \bibinfo {pages} {127004} (\bibinfo {year} {2007})}\BibitemShut {NoStop}%
\bibitem [{\citenamefont {Hasik}\ \emph
  {et~al.}(2021{\natexlab{b}})\citenamefont {Hasik}, \citenamefont
  {Poilblanc},\ and\ \citenamefont {Becca}}]{10.21468/SciPostPhys.10.1.012}%
  \BibitemOpen
  \bibfield  {author} {\bibinfo {author} {\bibfnamefont {J.}~\bibnamefont
  {Hasik}}, \bibinfo {author} {\bibfnamefont {D.}~\bibnamefont {Poilblanc}},\
  and\ \bibinfo {author} {\bibfnamefont {F.}~\bibnamefont {Becca}},\ }\bibfield
   {title} {\bibinfo {title} {{Investigation of the Néel phase of the
  frustrated Heisenberg antiferromagnet by differentiable symmetric tensor
  networks}},\ }\href {https://doi.org/10.21468/SciPostPhys.10.1.012}
  {\bibfield  {journal} {\bibinfo  {journal} {SciPost Phys.}\ }\textbf
  {\bibinfo {volume} {10}},\ \bibinfo {pages} {012} (\bibinfo {year}
  {2021}{\natexlab{b}})}\BibitemShut {NoStop}%
\bibitem [{\citenamefont {Poilblanc}(2017)}]{PhysRevB.96.121118}%
  \BibitemOpen
  \bibfield  {author} {\bibinfo {author} {\bibfnamefont {D.}~\bibnamefont
  {Poilblanc}},\ }\bibfield  {title} {\bibinfo {title} {Investigation of the
  chiral antiferromagnetic heisenberg model using projected entangled pair
  states},\ }\href {https://doi.org/10.1103/PhysRevB.96.121118} {\bibfield
  {journal} {\bibinfo  {journal} {Phys. Rev. B}\ }\textbf {\bibinfo {volume}
  {96}},\ \bibinfo {pages} {121118} (\bibinfo {year} {2017})}\BibitemShut
  {NoStop}%
\bibitem [{\citenamefont {Nishino}\ and\ \citenamefont
  {Okunishi}(1996)}]{nishino1996}%
  \BibitemOpen
  \bibfield  {author} {\bibinfo {author} {\bibfnamefont {T.}~\bibnamefont
  {Nishino}}\ and\ \bibinfo {author} {\bibfnamefont {K.}~\bibnamefont
  {Okunishi}},\ }\bibfield  {title} {\bibinfo {title} {{Corner Transfer Matrix
  Renormalization Group Method}},\ }\href {https://doi.org/10.1143/JPSJ.65.891}
  {\bibfield  {journal} {\bibinfo  {journal} {J. Phys. Soc. Jpn.}\ }\textbf
  {\bibinfo {volume} {65}},\ \bibinfo {pages} {891} (\bibinfo {year}
  {1996})}\BibitemShut {NoStop}%
\bibitem [{\citenamefont {Or\'us}\ and\ \citenamefont
  {Vidal}(2009)}]{orus2009-1}%
  \BibitemOpen
  \bibfield  {author} {\bibinfo {author} {\bibfnamefont {R.}~\bibnamefont
  {Or\'us}}\ and\ \bibinfo {author} {\bibfnamefont {G.}~\bibnamefont {Vidal}},\
  }\bibfield  {title} {\bibinfo {title} {Simulation of two-dimensional quantum
  systems on an infinite lattice revisited: Corner transfer matrix for tensor
  contraction},\ }\href {https://doi.org/10.1103/PhysRevB.80.094403} {\bibfield
   {journal} {\bibinfo  {journal} {Phys. Rev. B}\ }\textbf {\bibinfo {volume}
  {80}},\ \bibinfo {pages} {094403} (\bibinfo {year} {2009})}\BibitemShut
  {NoStop}%
\bibitem [{\citenamefont {Corboz}\ \emph {et~al.}(2014)\citenamefont {Corboz},
  \citenamefont {Rice},\ and\ \citenamefont {Troyer}}]{Corboz2014}%
  \BibitemOpen
  \bibfield  {author} {\bibinfo {author} {\bibfnamefont {P.}~\bibnamefont
  {Corboz}}, \bibinfo {author} {\bibfnamefont {T.~M.}\ \bibnamefont {Rice}},\
  and\ \bibinfo {author} {\bibfnamefont {M.}~\bibnamefont {Troyer}},\
  }\bibfield  {title} {\bibinfo {title} {Competing states in the $t$-${J}$
  model: Uniform $d$-wave state versus stripe state},\ }\href
  {https://doi.org/10.1103/PhysRevLett.113.046402} {\bibfield  {journal}
  {\bibinfo  {journal} {Phys. Rev. Lett.}\ }\textbf {\bibinfo {volume} {113}},\
  \bibinfo {pages} {046402} (\bibinfo {year} {2014})}\BibitemShut {NoStop}%
\bibitem [{\citenamefont {Liao}\ \emph {et~al.}(2019)\citenamefont {Liao},
  \citenamefont {Liu}, \citenamefont {Wang},\ and\ \citenamefont
  {Xiang}}]{Liao2017a}%
  \BibitemOpen
  \bibfield  {author} {\bibinfo {author} {\bibfnamefont {H.-J.}\ \bibnamefont
  {Liao}}, \bibinfo {author} {\bibfnamefont {J.-G.}\ \bibnamefont {Liu}},
  \bibinfo {author} {\bibfnamefont {L.}~\bibnamefont {Wang}},\ and\ \bibinfo
  {author} {\bibfnamefont {T.}~\bibnamefont {Xiang}},\ }\bibfield  {title}
  {\bibinfo {title} {Differentiable programming tensor networks},\ }\href
  {https://doi.org/10.1103/PhysRevX.9.031041} {\bibfield  {journal} {\bibinfo
  {journal} {Phys. Rev. X}\ }\textbf {\bibinfo {volume} {9}},\ \bibinfo {pages}
  {031041} (\bibinfo {year} {2019})}\BibitemShut {NoStop}%
\bibitem [{SM()}]{SM}%
  \BibitemOpen
  \href@noop {} {}\bibinfo {note} {See Supplemental Material which contains
  further details on the iPEPS simulations and their data
  analysis.}\BibitemShut {Stop}%
\bibitem [{\citenamefont {Huang}\ \emph {et~al.}(2024)\citenamefont {Huang},
  \citenamefont {Qian},\ and\ \citenamefont {Qin}}]{huang2023magnetization}%
  \BibitemOpen
  \bibfield  {author} {\bibinfo {author} {\bibfnamefont {J.}~\bibnamefont
  {Huang}}, \bibinfo {author} {\bibfnamefont {X.}~\bibnamefont {Qian}},\ and\
  \bibinfo {author} {\bibfnamefont {M.}~\bibnamefont {Qin}},\ }\bibfield
  {title} {\bibinfo {title} {On the magnetization of the 120° order of the
  spin-1/2 triangular lattice heisenberg model: a dmrg revisited},\ }\href
  {https://doi.org/10.1088/1361-648X/ad21a8} {\bibfield  {journal} {\bibinfo
  {journal} {Journal of Physics: Condensed Matter}\ }\textbf {\bibinfo {volume}
  {36}},\ \bibinfo {pages} {185602} (\bibinfo {year} {2024})}\BibitemShut
  {NoStop}%
\bibitem [{\citenamefont {Li}\ \emph {et~al.}(2022)\citenamefont {Li},
  \citenamefont {Li}, \citenamefont {Zhao}, \citenamefont {Luo},\ and\
  \citenamefont {Xie}}]{li2022magnetization}%
  \BibitemOpen
  \bibfield  {author} {\bibinfo {author} {\bibfnamefont {Q.}~\bibnamefont
  {Li}}, \bibinfo {author} {\bibfnamefont {H.}~\bibnamefont {Li}}, \bibinfo
  {author} {\bibfnamefont {J.}~\bibnamefont {Zhao}}, \bibinfo {author}
  {\bibfnamefont {H.-G.}\ \bibnamefont {Luo}},\ and\ \bibinfo {author}
  {\bibfnamefont {Z.~Y.}\ \bibnamefont {Xie}},\ }\bibfield  {title} {\bibinfo
  {title} {Magnetization of the spin-$\frac{1}{2}$ heisenberg antiferromagnet
  on the triangular lattice},\ }\href
  {https://doi.org/10.1103/PhysRevB.105.184418} {\bibfield  {journal} {\bibinfo
   {journal} {Phys. Rev. B}\ }\textbf {\bibinfo {volume} {105}},\ \bibinfo
  {pages} {184418} (\bibinfo {year} {2022})}\BibitemShut {NoStop}%
\bibitem [{\citenamefont {Chi}\ \emph {et~al.}(2022)\citenamefont {Chi},
  \citenamefont {Liu}, \citenamefont {Wan}, \citenamefont {Liao},\ and\
  \citenamefont {Xiang}}]{PhysRevLett.129.227201}%
  \BibitemOpen
  \bibfield  {author} {\bibinfo {author} {\bibfnamefont {R.}~\bibnamefont
  {Chi}}, \bibinfo {author} {\bibfnamefont {Y.}~\bibnamefont {Liu}}, \bibinfo
  {author} {\bibfnamefont {Y.}~\bibnamefont {Wan}}, \bibinfo {author}
  {\bibfnamefont {H.-J.}\ \bibnamefont {Liao}},\ and\ \bibinfo {author}
  {\bibfnamefont {T.}~\bibnamefont {Xiang}},\ }\bibfield  {title} {\bibinfo
  {title} {Spin excitation spectra of anisotropic spin-$1/2$ triangular lattice
  heisenberg antiferromagnets},\ }\href
  {https://doi.org/10.1103/PhysRevLett.129.227201} {\bibfield  {journal}
  {\bibinfo  {journal} {Phys. Rev. Lett.}\ }\textbf {\bibinfo {volume} {129}},\
  \bibinfo {pages} {227201} (\bibinfo {year} {2022})}\BibitemShut {NoStop}%
\bibitem [{\citenamefont {Iqbal}\ \emph {et~al.}(2016)\citenamefont {Iqbal},
  \citenamefont {Hu}, \citenamefont {Thomale}, \citenamefont {Poilblanc},\ and\
  \citenamefont {Becca}}]{PhysRevB.93.144411}%
  \BibitemOpen
  \bibfield  {author} {\bibinfo {author} {\bibfnamefont {Y.}~\bibnamefont
  {Iqbal}}, \bibinfo {author} {\bibfnamefont {W.-J.}\ \bibnamefont {Hu}},
  \bibinfo {author} {\bibfnamefont {R.}~\bibnamefont {Thomale}}, \bibinfo
  {author} {\bibfnamefont {D.}~\bibnamefont {Poilblanc}},\ and\ \bibinfo
  {author} {\bibfnamefont {F.}~\bibnamefont {Becca}},\ }\bibfield  {title}
  {\bibinfo {title} {Spin liquid nature in the heisenberg
  ${J}_{1}\ensuremath{-}{J}_{2}$ triangular antiferromagnet},\ }\href
  {https://doi.org/10.1103/PhysRevB.93.144411} {\bibfield  {journal} {\bibinfo
  {journal} {Phys. Rev. B}\ }\textbf {\bibinfo {volume} {93}},\ \bibinfo
  {pages} {144411} (\bibinfo {year} {2016})}\BibitemShut {NoStop}%
\bibitem [{\citenamefont {Kaneko}\ \emph {et~al.}(2014)\citenamefont {Kaneko},
  \citenamefont {Morita},\ and\ \citenamefont
  {Imada}}]{doi:10.7566/JPSJ.83.093707}%
  \BibitemOpen
  \bibfield  {author} {\bibinfo {author} {\bibfnamefont {R.}~\bibnamefont
  {Kaneko}}, \bibinfo {author} {\bibfnamefont {S.}~\bibnamefont {Morita}},\
  and\ \bibinfo {author} {\bibfnamefont {M.}~\bibnamefont {Imada}},\ }\bibfield
   {title} {\bibinfo {title} {Gapless spin-liquid phase in an extended spin 1/2
  triangular heisenberg model},\ }\href
  {https://doi.org/10.7566/JPSJ.83.093707} {\bibfield  {journal} {\bibinfo
  {journal} {Journal of the Physical Society of Japan}\ }\textbf {\bibinfo
  {volume} {83}},\ \bibinfo {pages} {093707} (\bibinfo {year}
  {2014})}\BibitemShut {NoStop}%
\bibitem [{\citenamefont {Niesen}\ and\ \citenamefont
  {Corboz}(2018)}]{niesen2018}%
  \BibitemOpen
  \bibfield  {author} {\bibinfo {author} {\bibfnamefont {I.}~\bibnamefont
  {Niesen}}\ and\ \bibinfo {author} {\bibfnamefont {P.}~\bibnamefont
  {Corboz}},\ }\bibfield  {title} {\bibinfo {title} {Ground-state study of the
  spin-1 bilinear-biquadratic heisenberg model on the triangular lattice using
  tensor networks},\ }\href {https://doi.org/10.1103/PhysRevB.97.245146}
  {\bibfield  {journal} {\bibinfo  {journal} {Phys. Rev. B}\ }\textbf {\bibinfo
  {volume} {97}},\ \bibinfo {pages} {245146} (\bibinfo {year}
  {2018})}\BibitemShut {NoStop}%
\bibitem [{\citenamefont {{Corboz}}\ \emph {et~al.}(2018)\citenamefont
  {{Corboz}}, \citenamefont {{Czarnik}}, \citenamefont {{Kapteijns}},\ and\
  \citenamefont {{Tagliacozzo}}}]{Corboz2018}%
  \BibitemOpen
  \bibfield  {author} {\bibinfo {author} {\bibfnamefont {P.}~\bibnamefont
  {{Corboz}}}, \bibinfo {author} {\bibfnamefont {P.}~\bibnamefont {{Czarnik}}},
  \bibinfo {author} {\bibfnamefont {G.}~\bibnamefont {{Kapteijns}}},\ and\
  \bibinfo {author} {\bibfnamefont {L.}~\bibnamefont {{Tagliacozzo}}},\
  }\bibfield  {title} {\bibinfo {title} {{Finite Correlation Length Scaling
  with Infinite Projected Entangled-Pair States}},\ }\href
  {https://doi.org/10.1103/PhysRevX.8.031031} {\bibfield  {journal} {\bibinfo
  {journal} {Phys. Rev. X}\ }\textbf {\bibinfo {volume} {8}},\ \bibinfo {pages}
  {031031} (\bibinfo {year} {2018})}\BibitemShut {NoStop}%
\bibitem [{\citenamefont {{Rader}}\ and\ \citenamefont
  {{L{\"a}uchli}}(2018)}]{Rader2018}%
  \BibitemOpen
  \bibfield  {author} {\bibinfo {author} {\bibfnamefont {M.}~\bibnamefont
  {{Rader}}}\ and\ \bibinfo {author} {\bibfnamefont {A.~M.}\ \bibnamefont
  {{L{\"a}uchli}}},\ }\bibfield  {title} {\bibinfo {title} {{Finite Correlation
  Length Scaling in Lorentz-Invariant Gapless iPEPS Wave Functions}},\ }\href
  {https://doi.org/10.1103/PhysRevX.8.031030} {\bibfield  {journal} {\bibinfo
  {journal} {Phys. Rev. X}\ }\textbf {\bibinfo {volume} {8}},\ \bibinfo {pages}
  {031030} (\bibinfo {year} {2018})}\BibitemShut {NoStop}%
\bibitem [{\citenamefont {Hasenfratz}\ and\ \citenamefont
  {Niedermayer}(1993)}]{hasenfratz93}%
  \BibitemOpen
  \bibfield  {author} {\bibinfo {author} {\bibfnamefont {P.}~\bibnamefont
  {Hasenfratz}}\ and\ \bibinfo {author} {\bibfnamefont {F.}~\bibnamefont
  {Niedermayer}},\ }\bibfield  {title} {{\selectlanguage {english}\bibinfo
  {title} {Finite size and temperature effects in the {AF} {Heisenberg}
  model}},\ }\href {https://doi.org/10.1007/BF01309171} {\bibfield  {journal}
  {\bibinfo  {journal} {Z. Physik B - Condensed Matter}\ }\textbf {\bibinfo
  {volume} {92}},\ \bibinfo {pages} {91} (\bibinfo {year} {1993})}\BibitemShut
  {NoStop}%
\bibitem [{\citenamefont {T\'oth}\ \emph {et~al.}(2010)\citenamefont {T\'oth},
  \citenamefont {L\"auchli}, \citenamefont {Mila},\ and\ \citenamefont
  {Penc}}]{toth2010}%
  \BibitemOpen
  \bibfield  {author} {\bibinfo {author} {\bibfnamefont {T.~A.}\ \bibnamefont
  {T\'oth}}, \bibinfo {author} {\bibfnamefont {A.~M.}\ \bibnamefont
  {L\"auchli}}, \bibinfo {author} {\bibfnamefont {F.}~\bibnamefont {Mila}},\
  and\ \bibinfo {author} {\bibfnamefont {K.}~\bibnamefont {Penc}},\ }\bibfield
  {title} {\bibinfo {title} {{Three-Sublattice Ordering of the SU(3) Heisenberg
  Model of Three-Flavor Fermions on the Square and Cubic Lattices}},\ }\href
  {https://doi.org/10.1103/PhysRevLett.105.265301} {\bibfield  {journal}
  {\bibinfo  {journal} {Phys. Rev. Lett.}\ }\textbf {\bibinfo {volume} {105}},\
  \bibinfo {pages} {265301} (\bibinfo {year} {2010})}\BibitemShut {NoStop}%
\bibitem [{\citenamefont {Bauer}\ \emph {et~al.}(2012)\citenamefont {Bauer},
  \citenamefont {Corboz}, \citenamefont {L\"{a}uchli}, \citenamefont {Messio},
  \citenamefont {Penc}, \citenamefont {Troyer},\ and\ \citenamefont
  {Mila}}]{bauer2011-su3}%
  \BibitemOpen
  \bibfield  {author} {\bibinfo {author} {\bibfnamefont {B.}~\bibnamefont
  {Bauer}}, \bibinfo {author} {\bibfnamefont {P.}~\bibnamefont {Corboz}},
  \bibinfo {author} {\bibfnamefont {A.~M.}\ \bibnamefont {L\"{a}uchli}},
  \bibinfo {author} {\bibfnamefont {L.}~\bibnamefont {Messio}}, \bibinfo
  {author} {\bibfnamefont {K.}~\bibnamefont {Penc}}, \bibinfo {author}
  {\bibfnamefont {M.}~\bibnamefont {Troyer}},\ and\ \bibinfo {author}
  {\bibfnamefont {F.}~\bibnamefont {Mila}},\ }\bibfield  {title} {\bibinfo
  {title} {Three-sublattice order in the {SU(3)} heisenberg model on the square
  and triangular lattice},\ }\href {https://doi.org/10.1103/PhysRevB.85.125116}
  {\bibfield  {journal} {\bibinfo  {journal} {Phys. Rev. B}\ }\textbf {\bibinfo
  {volume} {85}},\ \bibinfo {pages} {125116} (\bibinfo {year}
  {2012})}\BibitemShut {NoStop}%
\bibitem [{\citenamefont {Hasik}\ and\ \citenamefont {Mbeng}()}]{pepstorch}%
  \BibitemOpen
  \bibfield  {author} {\bibinfo {author} {\bibfnamefont {J.}~\bibnamefont
  {Hasik}}\ and\ \bibinfo {author} {\bibfnamefont {G.~B.}\ \bibnamefont
  {Mbeng}},\ }\href {https://github.com/jurajHasik/peps-torch} {\bibinfo
  {title} {{peps-torch: A differentiable tensor network library for
  two-dimensional lattice models}}},\ \bibinfo {howpublished}
  {\texttt{https://github.com/jurajHasik/peps-torch}}\BibitemShut {NoStop}%
\bibitem [{\citenamefont {Rams}\ \emph {et~al.}()\citenamefont {Rams},
  \citenamefont {Wójtowicz}, \citenamefont {Sinha},\ and\ \citenamefont
  {Hasik}}]{yastn}%
  \BibitemOpen
  \bibfield  {author} {\bibinfo {author} {\bibfnamefont {M.~M.}\ \bibnamefont
  {Rams}}, \bibinfo {author} {\bibfnamefont {G.}~\bibnamefont {Wójtowicz}},
  \bibinfo {author} {\bibfnamefont {A.}~\bibnamefont {Sinha}},\ and\ \bibinfo
  {author} {\bibfnamefont {J.}~\bibnamefont {Hasik}},\ }\href
  {https://github.com/yastn/yastn} {\bibinfo {title} {{Python library for
  differentiable linear algebra with block-sparse tensors, supporting abelian
  symmetries}}},\ \bibinfo {howpublished}
  {\texttt{https://github.com/yastn/yastn}}\BibitemShut {NoStop}%
\end{thebibliography}%


\begin{thebibliography}{8}%
\makeatletter
\providecommand \@ifxundefined [1]{%
 \@ifx{#1\undefined}
}%
\providecommand \@ifnum [1]{%
 \ifnum #1\expandafter \@firstoftwo
 \else \expandafter \@secondoftwo
 \fi
}%
\providecommand \@ifx [1]{%
 \ifx #1\expandafter \@firstoftwo
 \else \expandafter \@secondoftwo
 \fi
}%
\providecommand \natexlab [1]{#1}%
\providecommand \enquote  [1]{``#1''}%
\providecommand \bibnamefont  [1]{#1}%
\providecommand \bibfnamefont [1]{#1}%
\providecommand \citenamefont [1]{#1}%
\providecommand \href@noop [0]{\@secondoftwo}%
\providecommand \href [0]{\begingroup \@sanitize@url \@href}%
\providecommand \@href[1]{\@@startlink{#1}\@@href}%
\providecommand \@@href[1]{\endgroup#1\@@endlink}%
\providecommand \@sanitize@url [0]{\catcode `\\12\catcode `\$12\catcode
  `\&12\catcode `\#12\catcode `\^12\catcode `\_12\catcode `\%12\relax}%
\providecommand \@@startlink[1]{}%
\providecommand \@@endlink[0]{}%
\providecommand \url  [0]{\begingroup\@sanitize@url \@url }%
\providecommand \@url [1]{\endgroup\@href {#1}{\urlprefix }}%
\providecommand \urlprefix  [0]{URL }%
\providecommand \Eprint [0]{\href }%
\providecommand \doibase [0]{https://doi.org/}%
\providecommand \selectlanguage [0]{\@gobble}%
\providecommand \bibinfo  [0]{\@secondoftwo}%
\providecommand \bibfield  [0]{\@secondoftwo}%
\providecommand \translation [1]{[#1]}%
\providecommand \BibitemOpen [0]{}%
\providecommand \bibitemStop [0]{}%
\providecommand \bibitemNoStop [0]{.\EOS\space}%
\providecommand \EOS [0]{\spacefactor3000\relax}%
\providecommand \BibitemShut  [1]{\csname bibitem#1\endcsname}%
\let\auto@bib@innerbib\@empty
\bibitem [{\citenamefont {Corboz}\ \emph {et~al.}(2014)\citenamefont {Corboz},
  \citenamefont {Rice},\ and\ \citenamefont {Troyer}}]{corboz14_tJ}%
  \BibitemOpen
  \bibfield  {author} {\bibinfo {author} {\bibfnamefont {P.}~\bibnamefont
  {Corboz}}, \bibinfo {author} {\bibfnamefont {T.~M.}\ \bibnamefont {Rice}},\
  and\ \bibinfo {author} {\bibfnamefont {M.}~\bibnamefont {Troyer}},\
  }\bibfield  {title} {\bibinfo {title} {Competing {States} in the t-{J}
  {Model}: {Uniform} d-{Wave} {State} versus {Stripe} {State}},\ }\href
  {https://doi.org/10.1103/PhysRevLett.113.046402} {\bibfield  {journal}
  {\bibinfo  {journal} {Phys. Rev. Lett.}\ }\textbf {\bibinfo {volume} {113}},\
  \bibinfo {pages} {046402} (\bibinfo {year} {2014})}\BibitemShut {NoStop}%
\bibitem [{\citenamefont {Hasik}\ and\ \citenamefont {Mbeng}()}]{pepstorch}%
  \BibitemOpen
  \bibfield  {author} {\bibinfo {author} {\bibfnamefont {J.}~\bibnamefont
  {Hasik}}\ and\ \bibinfo {author} {\bibfnamefont {G.~B.}\ \bibnamefont
  {Mbeng}},\ }\href {https://github.com/jurajHasik/peps-torch} {\bibinfo
  {title} {{peps-torch: A differentiable tensor network library for
  two-dimensional lattice models}}},\ \bibinfo {howpublished}
  {\texttt{https://github.com/jurajHasik/peps-torch}}\BibitemShut {NoStop}%
\bibitem [{\citenamefont {Press}\ \emph {et~al.}(2007)\citenamefont {Press},
  \citenamefont {Teukolsky}, \citenamefont {Vetterling},\ and\ \citenamefont
  {Flannery}}]{Numerical2007}%
  \BibitemOpen
  \bibfield  {author} {\bibinfo {author} {\bibfnamefont {W.~H.}\ \bibnamefont
  {Press}}, \bibinfo {author} {\bibfnamefont {S.~A.}\ \bibnamefont
  {Teukolsky}}, \bibinfo {author} {\bibfnamefont {W.~P.}\ \bibnamefont
  {Vetterling}},\ and\ \bibinfo {author} {\bibfnamefont {B.~P.}\ \bibnamefont
  {Flannery}},\ }\href@noop {} {\emph {\bibinfo {title} {Numerical Recipes}}}\
  (\bibinfo  {publisher} {Cambridge University Press},\ \bibinfo {year}
  {2007})\BibitemShut {NoStop}%
\bibitem [{\citenamefont {Nishino}\ \emph {et~al.}(1996)\citenamefont
  {Nishino}, \citenamefont {Okunishi},\ and\ \citenamefont
  {Kikuchi}}]{NISHINO199669}%
  \BibitemOpen
  \bibfield  {author} {\bibinfo {author} {\bibfnamefont {T.}~\bibnamefont
  {Nishino}}, \bibinfo {author} {\bibfnamefont {K.}~\bibnamefont {Okunishi}},\
  and\ \bibinfo {author} {\bibfnamefont {M.}~\bibnamefont {Kikuchi}},\
  }\bibfield  {title} {\bibinfo {title} {Numerical renormalization group at
  criticality},\ }\href
  {https://doi.org/https://doi.org/10.1016/0375-9601(96)00128-4} {\bibfield
  {journal} {\bibinfo  {journal} {Physics Letters A}\ }\textbf {\bibinfo
  {volume} {213}},\ \bibinfo {pages} {69} (\bibinfo {year} {1996})}\BibitemShut
  {NoStop}%
\bibitem [{\citenamefont {Rams}\ \emph {et~al.}(2018)\citenamefont {Rams},
  \citenamefont {Czarnik},\ and\ \citenamefont {Cincio}}]{rams2018}%
  \BibitemOpen
  \bibfield  {author} {\bibinfo {author} {\bibfnamefont {M.~M.}\ \bibnamefont
  {Rams}}, \bibinfo {author} {\bibfnamefont {P.}~\bibnamefont {Czarnik}},\ and\
  \bibinfo {author} {\bibfnamefont {L.}~\bibnamefont {Cincio}},\ }\bibfield
  {title} {\bibinfo {title} {Precise extrapolation of the correlation function
  asymptotics in uniform tensor network states with application to the
  bose-hubbard and xxz models},\ }\href
  {https://doi.org/10.1103/PhysRevX.8.041033} {\bibfield  {journal} {\bibinfo
  {journal} {Phys. Rev. X}\ }\textbf {\bibinfo {volume} {8}},\ \bibinfo {pages}
  {041033} (\bibinfo {year} {2018})}\BibitemShut {NoStop}%
\bibitem [{\citenamefont {Hasik}\ \emph {et~al.}(2021)\citenamefont {Hasik},
  \citenamefont {Poilblanc},\ and\ \citenamefont
  {Becca}}]{10.21468/SciPostPhys.10.1.012}%
  \BibitemOpen
  \bibfield  {author} {\bibinfo {author} {\bibfnamefont {J.}~\bibnamefont
  {Hasik}}, \bibinfo {author} {\bibfnamefont {D.}~\bibnamefont {Poilblanc}},\
  and\ \bibinfo {author} {\bibfnamefont {F.}~\bibnamefont {Becca}},\ }\bibfield
   {title} {\bibinfo {title} {{Investigation of the Néel phase of the
  frustrated Heisenberg antiferromagnet by differentiable symmetric tensor
  networks}},\ }\href {https://doi.org/10.21468/SciPostPhys.10.1.012}
  {\bibfield  {journal} {\bibinfo  {journal} {SciPost Phys.}\ }\textbf
  {\bibinfo {volume} {10}},\ \bibinfo {pages} {012} (\bibinfo {year}
  {2021})}\BibitemShut {NoStop}%
\bibitem [{\citenamefont {Hasik}\ \emph {et~al.}(2022)\citenamefont {Hasik},
  \citenamefont {Mbeng}, \citenamefont {Capponi}, \citenamefont {Becca},\ and\
  \citenamefont {L\"auchli}}]{Hasik_2022}%
  \BibitemOpen
  \bibfield  {author} {\bibinfo {author} {\bibfnamefont {J.}~\bibnamefont
  {Hasik}}, \bibinfo {author} {\bibfnamefont {G.~B.}\ \bibnamefont {Mbeng}},
  \bibinfo {author} {\bibfnamefont {S.}~\bibnamefont {Capponi}}, \bibinfo
  {author} {\bibfnamefont {F.}~\bibnamefont {Becca}},\ and\ \bibinfo {author}
  {\bibfnamefont {A.~M.}\ \bibnamefont {L\"auchli}},\ }\bibfield  {title}
  {\bibinfo {title} {Symmetric projected entangled-pair states analysis of a
  phase transition in coupled spin-$\frac{1}{2}$ ladders},\ }\href
  {https://doi.org/10.1103/PhysRevB.106.125154} {\bibfield  {journal} {\bibinfo
   {journal} {Phys. Rev. B}\ }\textbf {\bibinfo {volume} {106}},\ \bibinfo
  {pages} {125154} (\bibinfo {year} {2022})}\BibitemShut {NoStop}%
\bibitem [{\citenamefont {Weihong}\ \emph {et~al.}(1999)\citenamefont
  {Weihong}, \citenamefont {McKenzie},\ and\ \citenamefont
  {Singh}}]{PhysRevB.59.14367}%
  \BibitemOpen
  \bibfield  {author} {\bibinfo {author} {\bibfnamefont {Z.}~\bibnamefont
  {Weihong}}, \bibinfo {author} {\bibfnamefont {R.~H.}\ \bibnamefont
  {McKenzie}},\ and\ \bibinfo {author} {\bibfnamefont {R.~R.~P.}\ \bibnamefont
  {Singh}},\ }\bibfield  {title} {\bibinfo {title} {Phase diagram for a class
  of spin-$\frac{1}{2}$ heisenberg models interpolating between the
  square-lattice, the triangular-lattice, and the linear-chain limits},\ }\href
  {https://doi.org/10.1103/PhysRevB.59.14367} {\bibfield  {journal} {\bibinfo
  {journal} {Phys. Rev. B}\ }\textbf {\bibinfo {volume} {59}},\ \bibinfo
  {pages} {14367} (\bibinfo {year} {1999})}\BibitemShut {NoStop}%
\end{thebibliography}%
    
    \end{document}


\title{Incommensurate order with translationally invariant projected entangled-pair states:
Spiral states and quantum spin liquid on the anisotropic triangular lattice:  \emph{Supplemental material}}

\author{Juraj Hasik}
\affiliation{Institute for Theoretical Physics, University of Amsterdam,
Science Park 904, 1098 XH Amsterdam, The Netherlands}

\author{Philippe Corboz}
\affiliation{Institute for Theoretical Physics, University of Amsterdam,
Science Park 904, 1098 XH Amsterdam, The Netherlands}

\maketitle

In this supplemental material we give further details on the iPEPS simulations and their data analysis. We outline the implementation of spiral iPEPS in Sec.~\ref{sec:spiral_extra}.
We define iPEPS correlation length in Sec.~\ref{sec:ipepstmmatrix} and  provide the detailed finite correlation-length scaling (FCLS) analysis of the local magnetizations for the $\mathbf{q}=(\pi,\pi)$ ordered U(1)-symmetric iPEPS in Sec.~\ref{sec:fclsu1} and for the spiral iPEPS in  Sec.~\ref{sec:fclspiral}, respectively. In Sec.~\ref{sec:vbs} we analyze the role of a competing dimerized state within the highly-frustrated region $J_{11}\in[0.73(1),0.8(2)]$.

\section{Spiral iPEPS and its variational optimization}
\label{sec:spiral_extra}

Here we describe in more detail the implementation of spiral iPEPS and its optimization.
Following the main text, we define the cost function as the energy per site $e(a,q;\chi)$ of the model. It is a function of the on-site tensor $a$ with $d D^4$ variational parameters, with local dimension $d=2$ for a spin S=1/2 system, and the spiral wave vector $\mathbf{q}=(2\pi q,2\pi q)$, parametrized by a single real variable $q$. Thus, in total we have $2D^4+1$ variational parameters, while the environment bond dimension $\chi$ is fixed. 


The evaluation of this cost function consists of three steps:
\begin{enumerate}
    \item Initialize the environment tensors, i.e. the four $\chi \times \chi$ corner tensors $\{C_1,\ldots,C_4\}$ and the four $\chi \times \chi \times D^2$ half-row/column transfer tensors $\{T_1,\ldots,T_4\}$. 
    This is typically done by random tensors or by contractions of bra and ket tensors $a$, i.e. $(C_1)_{(dd')(rr')}=\sum_{ul}a_{uldr}a^*_{uld'r'}$ or $(T_1)_{(ll')(dd')(rr')}=\sum_{u}a_{uldr}a^*_{uld'r'}$ and analogously for the other environment tensors.
    \item Perform directional CTMRG~\cite{corboz14_tJ} for single-site iPEPS generated by tensor $a$ until convergence. As convergence criterion, we use the 2-norm of the difference in singular value spectra of corner tensors between consecutive iterations, requiring this difference to be below at least $10^{-8}$. 
    \item Compute the energy per site by evaluating the spin-spin exchange terms along the three non-equivalent bonds, i.e in horizontal, vertical, and diagonal direction,  given by the contraction of the following tensor networks, respectively, 
        \begin{widetext}
            \begin{equation*}
                e(a,q;\chi)= J \vcenter{\hbox{\includegraphics[scale=0.2]{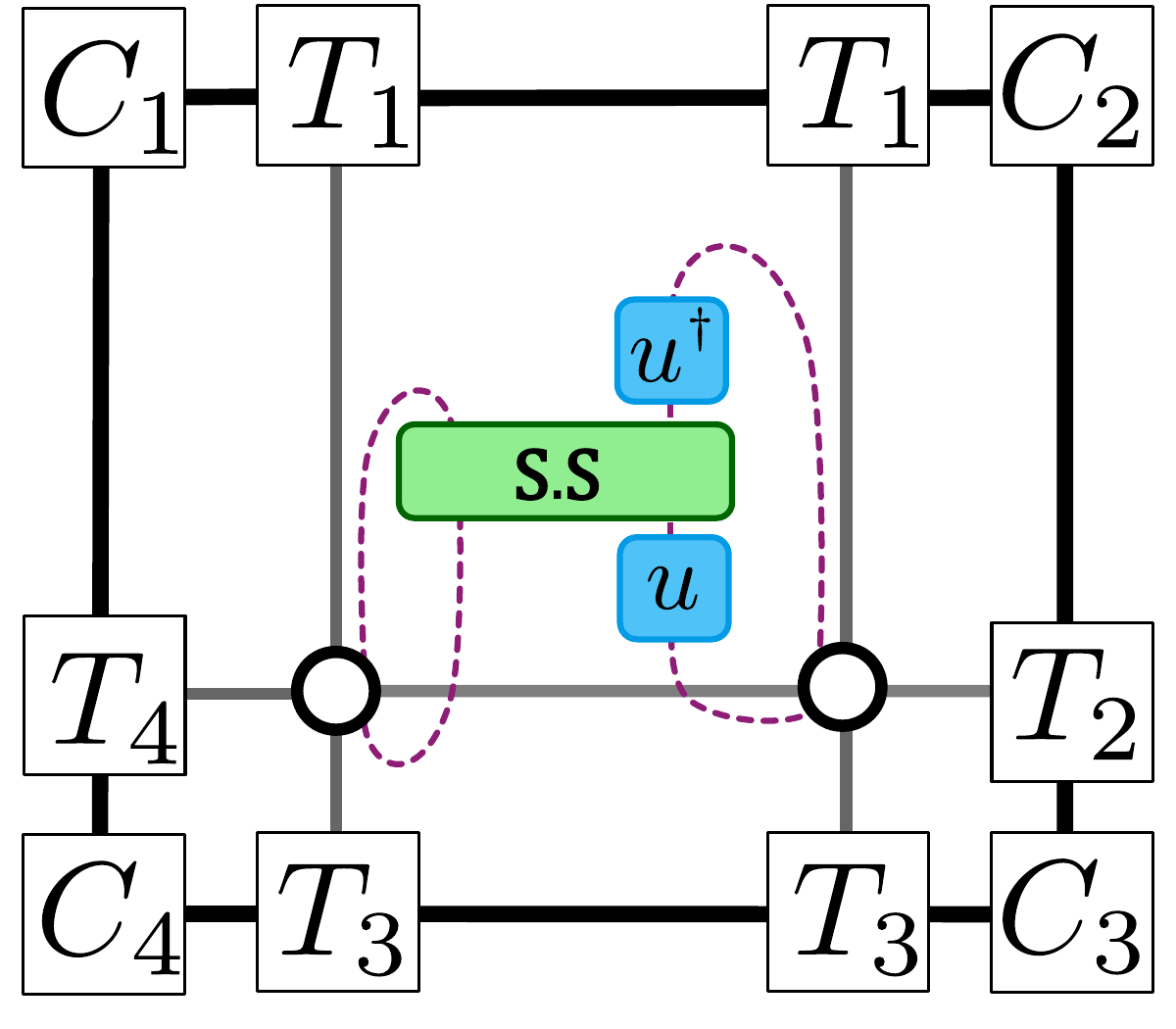}}} + J \vcenter{\hbox{\includegraphics[scale=0.2]{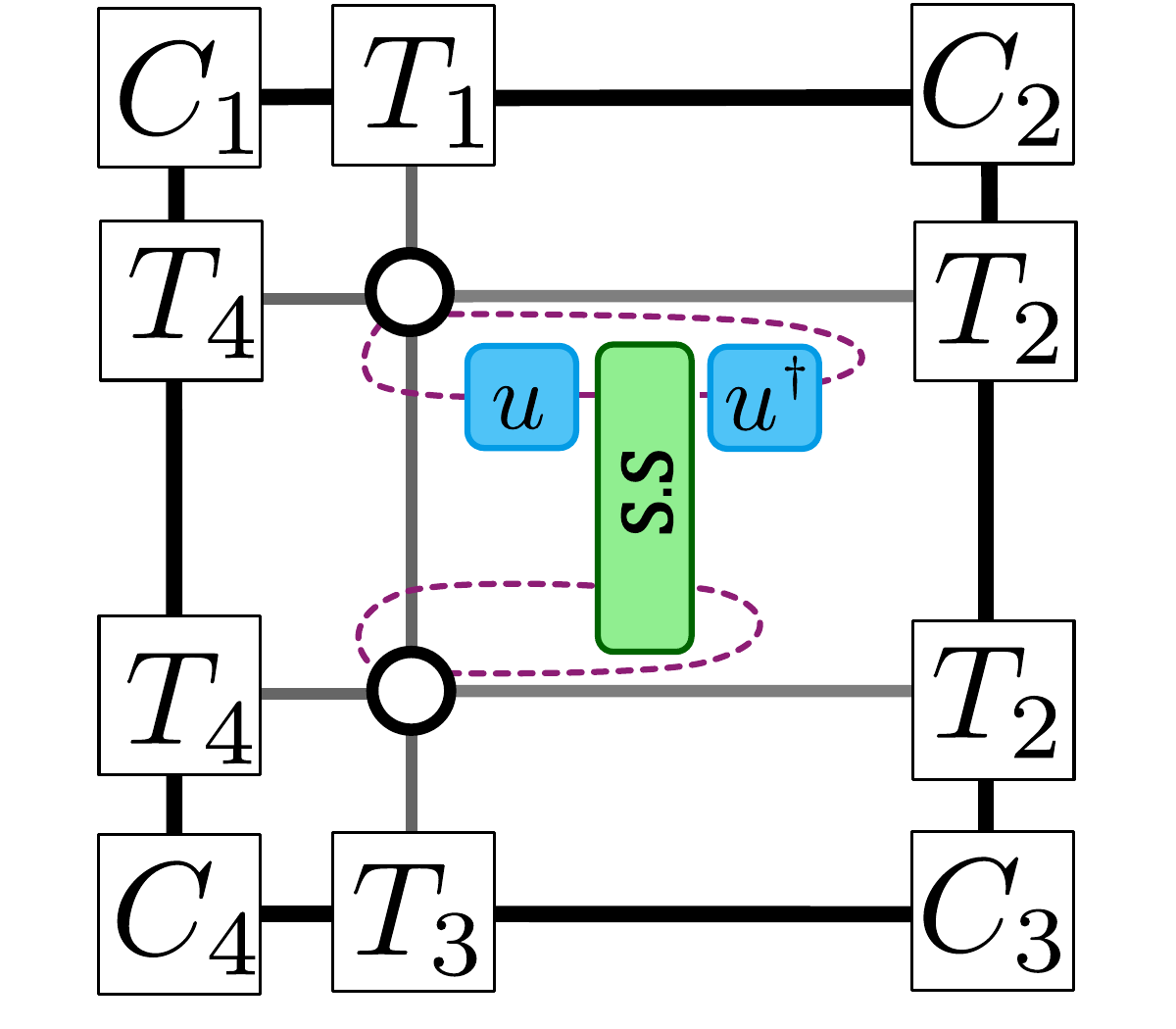}}} + J_{11} \vcenter{\hbox{\includegraphics[scale=0.2]{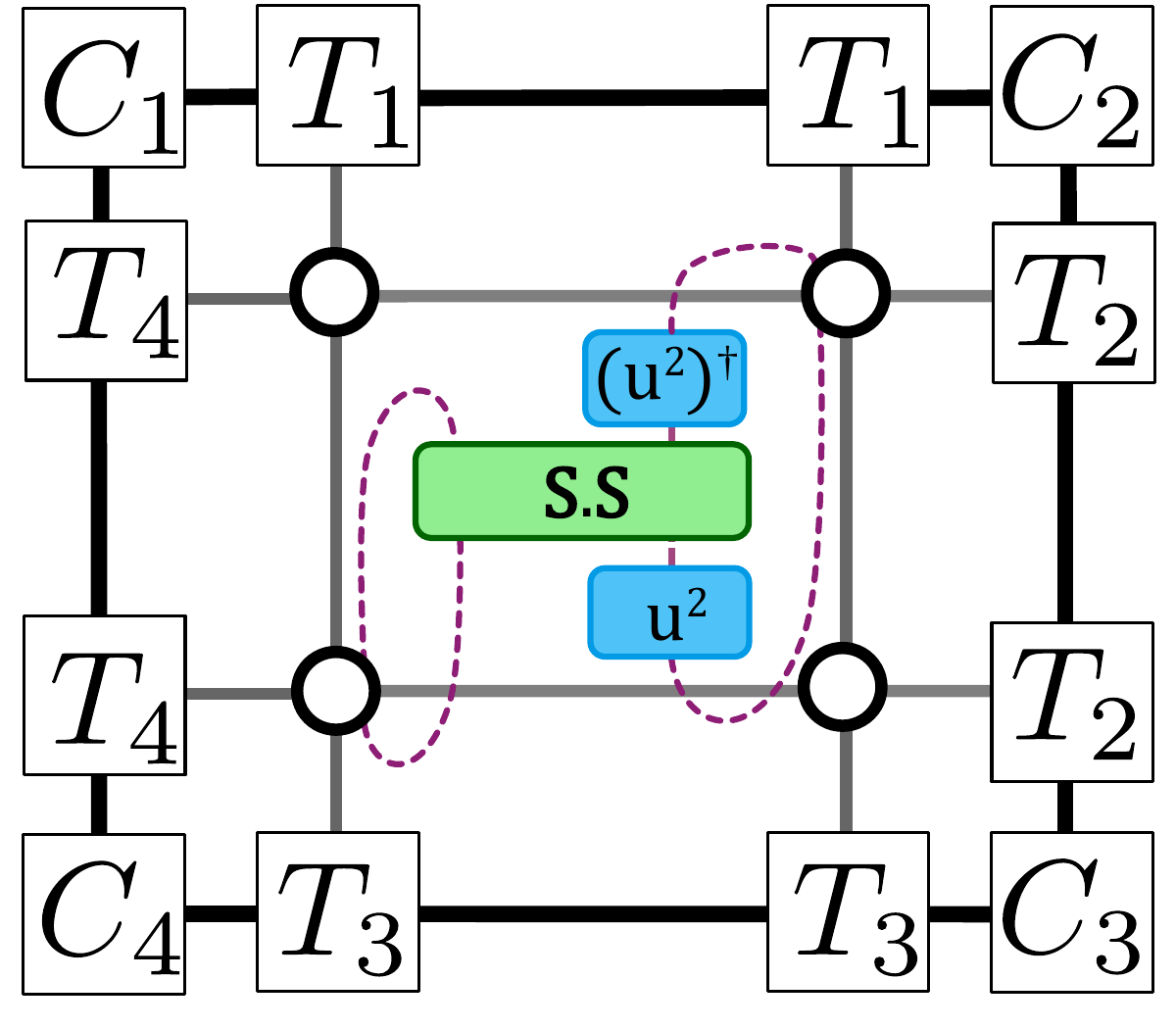}}},
            \end{equation*}
        \end{widetext}
        where the violet dashed lines represent physical indices of on-site tensors. These networks are the only place where the unitary $U(q)$, or rather the individual on-site unitaries $u_\mathbf{r}(\mathbf{q},\mathbf{r'})$, 
        denoted by blue boxes, appear explicitly. Here, we need to consider just a single $u(q,\Delta=1)=exp( i \pi q \sigma^y) $ with Pauli matrix $\sigma^y$. 
\end{enumerate}
We have implemented this cost function as a part of peps-torch~\cite{pepstorch}, a python libary based on PyTorch machine-learning framework. 
Working at fixed-$\chi$, which governs the precision of environment tensors, we evaluate $e(a,q;\chi)$ and its gradient with respect to both $a$ and $q$ by automatic differentiation.
Hence, the variational optimization of spiral iPEPS is now expressed as non-linear minimization and we solve it by gradient descent.
In particular, we use the L-BFGS algorithm with backtracking line search. In each optimization step, the values of the tensor $a$ as well as spiral's wave vector $q$ are updated.
We typically stop the optimization once the energy improvement from a optimization step drops below $10^{-7}$.
For further details regarding L-BFGS gradient descent, which is a quasi-Newton method constructing an approximate Hessian from past gradients, 
and line search, see Ref.~\cite{Numerical2007}. For fixed $D$, we typically start the optimizations at $\chi\gtrsim D^2$ and then further refine the optimization by increasing $\chi$ up to $4D^2$.

\section{iPEPS transfer matrix and correlation lengths}
\label{sec:ipepstmmatrix}

The iPEPS correlation functions and their correlation lengths are determined by the transfer matrix formed by an infinite column of contracted double-layer tensors, or its finite-$\chi$ approximation based on CTMRG, 
\begin{equation}
    \langle \mathcal{O}_0\mathcal{O}_{r+1}\rangle= \vcenter{\hbox{\includegraphics[scale=0.15]{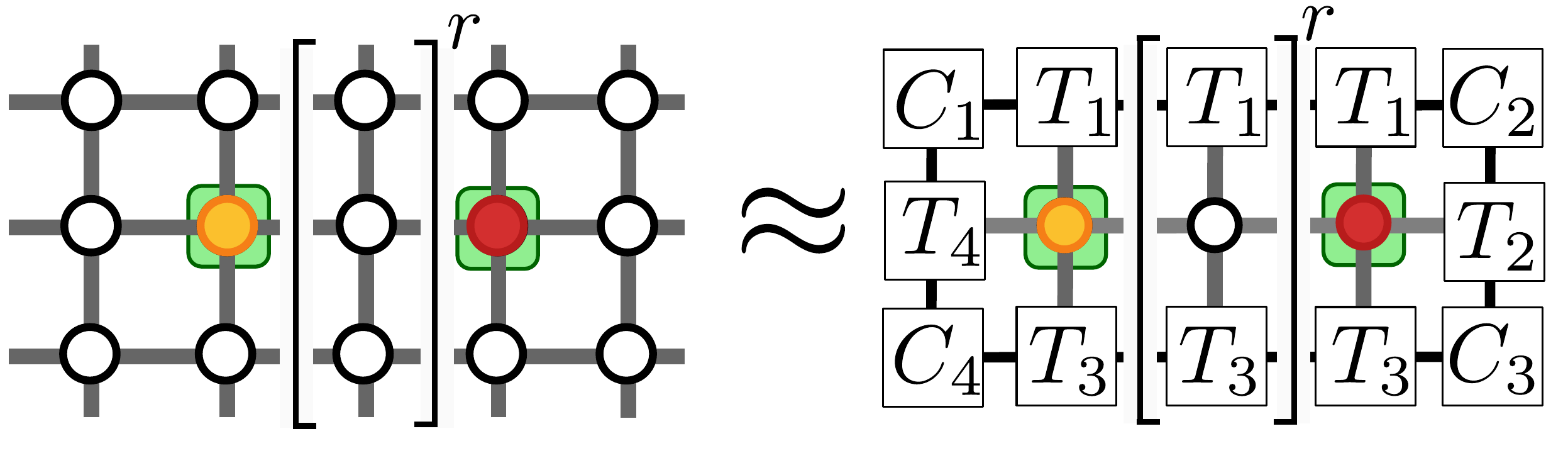}}},
\label{eq:pepstm}
\end{equation}
where the colored on-site site tensors account for observables $\mathcal{O}_0$ and $\mathcal{O}_{r+1}$ together with respective local unitaries $u_\mathbf{r}(\mathbf{q},\mathbf{r})$.
We compute the correlation length $\xi(\chi)$ from the normalized spectrum $\{1,\lambda_1,\lambda_2,\ldots\}$ of the $\chi^2 D^2 \times \chi^2 D^2$ transfer matrix~\cite{NISHINO199669}, here in square brackets, as $\xi(\chi)=-1/\textrm{log}\left[\lambda_1(\chi)\right]$ and then obtain its $\xi(\chi\to\infty)$ limit with the extrapolation using the second gap in the spectrum~\cite{rams2018}.
For the states analyzed here, $\xi$ describes the decay of the connected spin-spin correlations in the long distance limit.

\section{FCLS of the U(1) iPEPS data}
\label{sec:fclsu1}

To simulate the N\'eel phase, we take advantage of the U(1)-symmetry with respect to the magnetization axis and work with an iPEPS ansatz that is generated by a single U(1)-symmetric tensor $a$, with the optimal U(1) sectors taken from Ref.~\cite{10.21468/SciPostPhys.10.1.012}. 
The iPEPS tiles the lattice with a bipartite pattern, as shown in the Eq.~2 in the main text, where tensors on the b-sublattice are now obtained from tensor $a$ by U(1) charge conjugation. Such iPEPS can be consistently defined when the charge sectors on four virtual indices of the tensor are equivalent.
Beyond imposing the U(1) symmetry on physical observables, this construction also prevents any valence bond formation. After optimization, they constitute a single family of finite-$D$ variational minima amenable to extrapolations. 

Let us remark that energies evaluated at (relatively) modest finite environment dimensions $\chi\leq4D^2$, where optimizations are performed, might be lower than the variational value obtained in the $\chi\to\infty$ limit. This behavior, observed before in an iPEPS study of a spin-1/2 coupled ladders system in Ref.~\cite{Hasik_2022} and analogous to over-training in neural network context, stems from optimization of the finite-$\chi$ approximation of the energy rather than directly its variational estimate in the infinite $\chi$ limit. 
In the most severe cases,
for $D=7$ in the vicinity of the transition from the N\'eel to the QSL phase at $J_{11}\sim0.7$, this difference can be as large as $7\times10^{-5}$. It can be systematically suppressed by increasing $\chi$ at which the optimization is performed.

\begin{figure}[tb]
    \centering
    \includegraphics[width=\columnwidth]{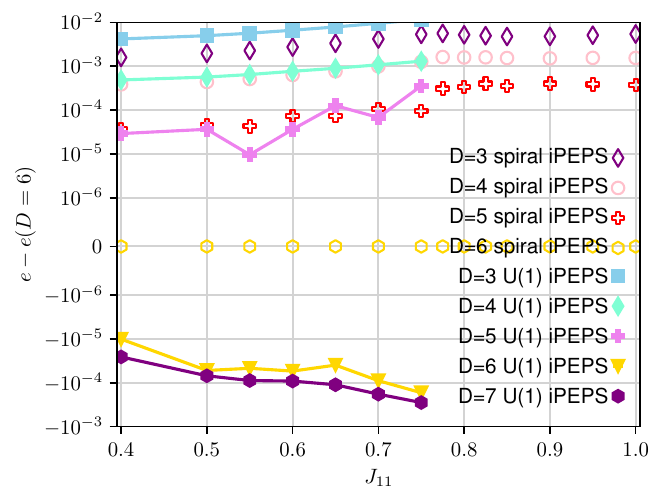}
    \caption{Comparison of energies per site between spiral iPEPS and U(1)-symmetric iPEPS, with the $D=6$ spiral iPEPS energy taken as the reference value.
   }
    \label{fig:u1_datae} 
\end{figure}
\begin{figure}[tb]
    \centering
    \includegraphics[width=\columnwidth]{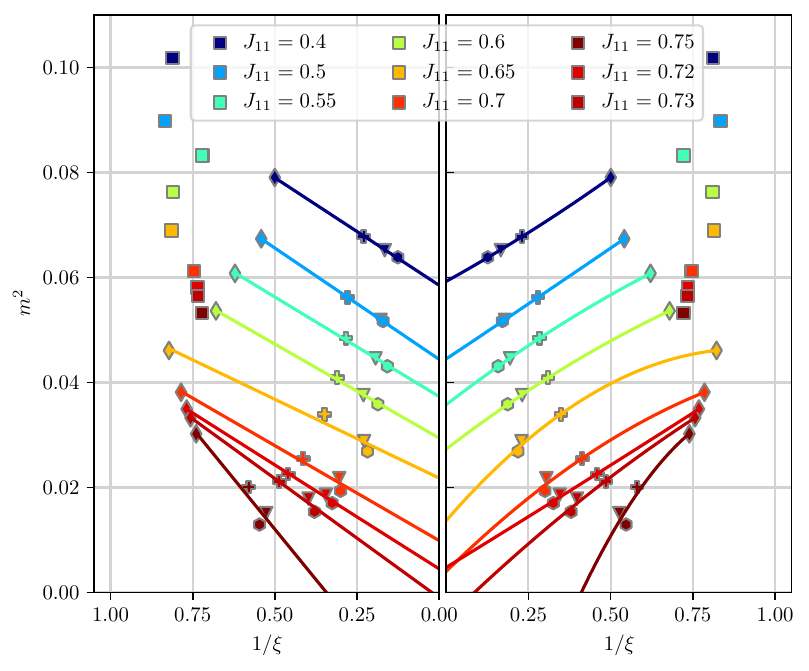}
    \caption{FCLS analysis of the U(1) iPEPS data. 
    Comparison of linear  and quadratic fit of $m^2$ as a function of correlation length, based on the $D\ge4$ data.
    The data for $D=3, \ldots, 7$ is given by the squares, diamonds, pluses, triangles, and hexagons, respectively. 
    }
    \label{fig:u1_datam} 
\end{figure}

In Fig.~\ref{fig:u1_datae} we present the energy of the  U(1)-symmetric iPEPS compared to the spiral iPEPS, showing that the N\'eel-ordered states  provide the best ground states  up $J_{11}\leq 0.75$ (at finite $D$). Beyond $J_{11}\geq0.8$, where the system enters the spiral phase with incommensurate $\mathbf{q}$, the energy of the U(1) iPEPS ansatz becomes uncompetitive. 

In Fig.~\ref{fig:u1_datam} we show the FCLS extrapolations for $m^2$ for a range of $J_{11}$ values, comparing linear and quadratic fits of the $D>3$ data. These two fits agree with a high precision up to $J_{11}\sim0.6$. The magnetic order vanishes between $J_{11}\in[0.72, 0.73]$ according to both fits. 
In general, as the order becomes weaker, the finite-$D$ effects  become more pronounced and higher-$D$ data are necessary to enter the regime where higher-order corrections to the scaling formula are negligible.

\begin{figure}[]
    \includegraphics[width=\columnwidth]{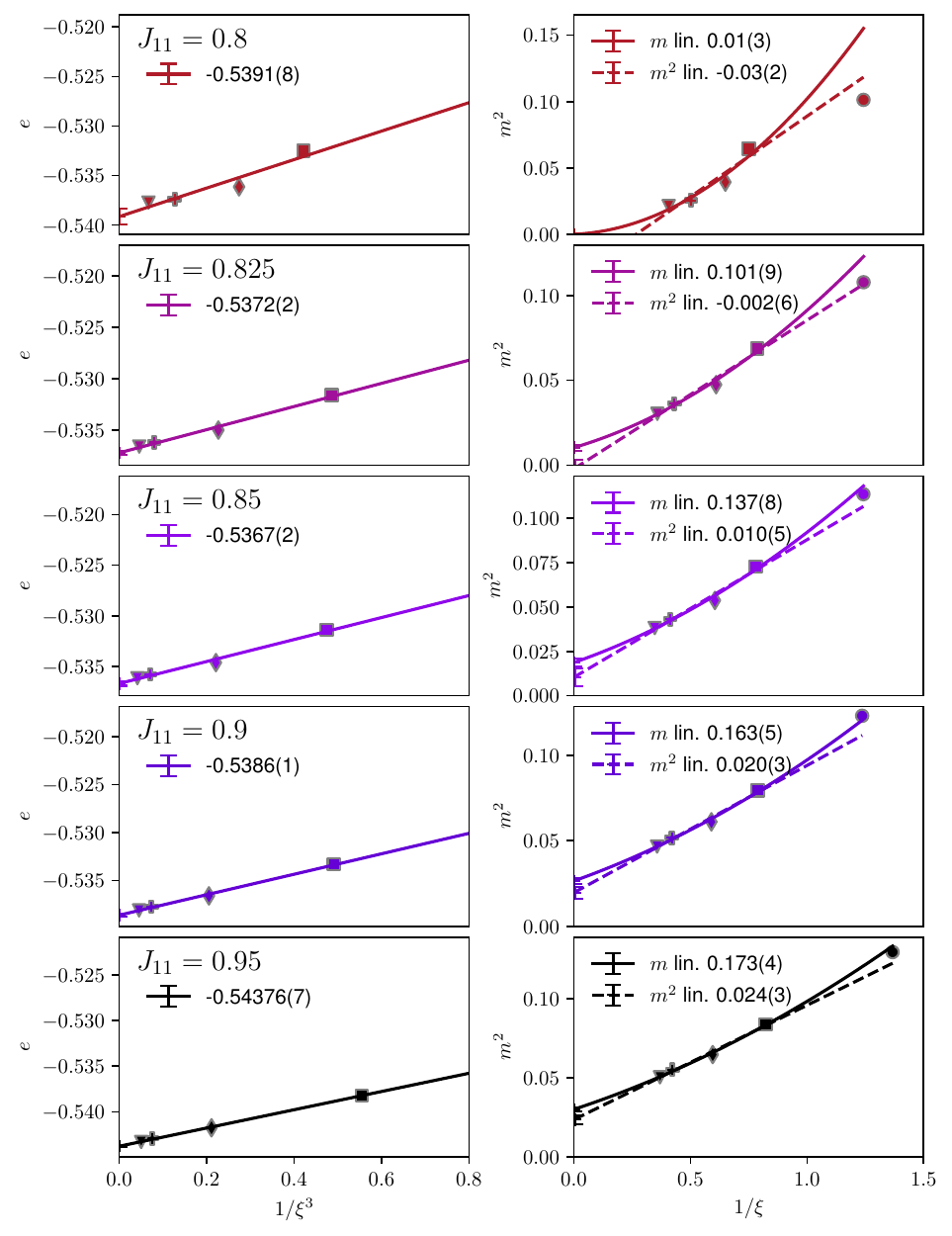}
    \caption{FCLS of the spiral iPEPS data. 
    Left panels: FCLS extrapolations of the energies based on a linear fit in $1/\xi^3$.
    Right panels: Comparison of extrapolations based on linear fits to $m$ and $m^2$.  The spiral iPEPS data for $D=2, \dots, 6$ are labeled by circles, squares, diamonds, pluses, and triangles, respectively. The $D=2$ data is excluded from the fits.
    }
    \label{fig:supp_spiralipeps} 
\end{figure}
\section{FCLS of the spiral iPEPS data}
\label{sec:fclspiral}

In this section, we give a detailed account of the FCLS extrapolations of the energy and local magnetization from the spiral iPEPS ansatz in the spiral phase $J_{11}\in[0.8,1.0]$. These results are shown in Fig.~\ref{fig:supp_spiralipeps}, except the case of $J_{11}=1$ which is discussed in the main text already. Remarkably, both the energy and local magnetization $m$ can be well captured by FCLS of the spiral iPEPS data. The difference in results between linear extrapolation of $m^2$, theoretically underpinned for the $\mathbf{q}=(\pi,\pi)$ order, and linear extrapolation of $m$ becomes significant only in the vicinity 
of the transition into the QSL phase around $J_{11}\sim0.8$.

\begin{figure}[]
    \centering
    \includegraphics[width=\columnwidth]{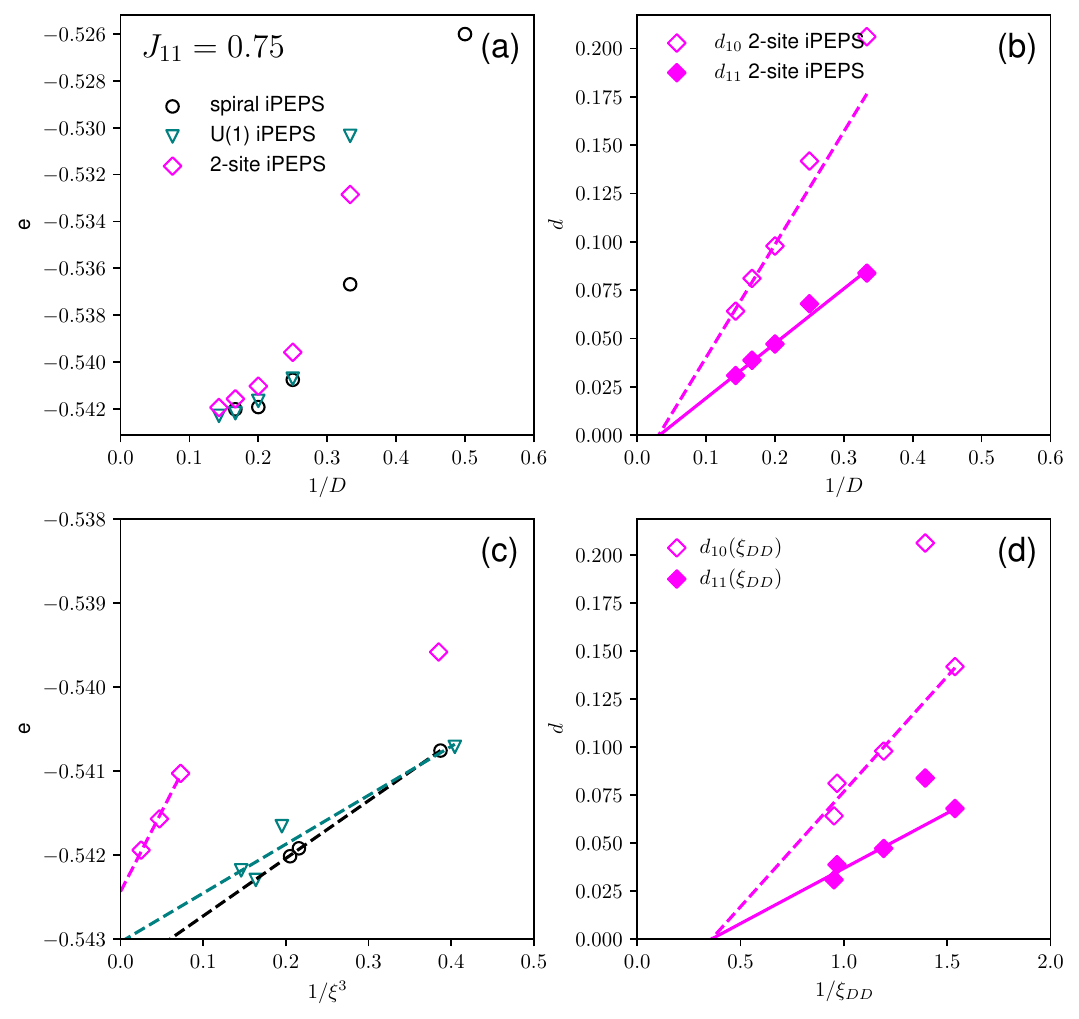}
    \caption{Results on the competing columnar-dimer state.
    (a)~Comparison of the energies per site between the 2-site iPEPS, spiral iPEPS, and U(1)-symmetric iPEPS.
    (b) Dimer order parameters $d_{10}$, $d_{11}$ of the 2-site iPEPS as a function of inverse bond dimension.
    (c)~Same as (a), but plotted as a function of the inverse third power of the correlation length.
    (d) Same as (b), but plotted against the dimer-dimer correlation length $\xi_{DD}$, extracted from fits to horizontal dimer-dimer correlation function.
    The (dashed) lines serve as guide to the eye.
    }
    \label{fig:supp_1sitevs2site} 
\end{figure}

\section{Valence bond solid candidate}
\label{sec:vbs}
In the final section of the supplemental material we address the role of the competing columnar-dimer state predicted by series expansion~\cite{PhysRevB.59.14367}
in the highly frustrated region $J_{11}\in[0.73(1),0.80(2)]$. We use an iPEPS ansatz with a $2\times1$ unit cell, featuring two independent tensors $a$ and $b$, which can naturally form dimer order and frustrates all spiral orders. The optimized 2-site iPEPS display vanishing local magnetization, with values of at most $m\sim O(10^{-4})$. In Fig.~\ref{fig:supp_1sitevs2site} we show the analysis of finite-$D$ and extrapolated data of the competing states at $J_{11}=0.75$ point, deep in the frustrated region.
The energetics show the $\mathbf{q}=(\pi,\pi)$ order to prevail at finite-$D$ (Fig.~\ref{fig:supp_1sitevs2site}(a)) which is further supported by the FCLS extrapolation of the energies displayed in  Fig.~\ref{fig:supp_1sitevs2site}(c).
We also analyze the dimer order parameters
\begin{align}
    d_{10}&=\langle \mathbf{S}_0\cdot\mathbf{S}_{\hat{x}}\rangle - \langle\mathbf{S}_{\hat{x}}\cdot\mathbf{S}_{2\hat{x}}\rangle,\\
    d_{11}&=\langle \mathbf{S}_0\cdot\mathbf{S}_{\hat{x}+\hat{y}}\rangle -\langle \mathbf{S}_{\hat{x}+\hat{y}}\cdot\mathbf{S}_{2\hat{x}+2\hat{y}}\rangle,
\end{align}
in horizontal and diagonal directions respectively.

Both extrapolations of the dimer order parameters in the inverse bond dimension and in the inverse correlation length shown in Figs.~\ref{fig:supp_1sitevs2site}(b) and (d) indicate that the dimer order vanishes in the infinite $D$ limit, providing further evidence that the columnar-dimer state predicted by series expansion is not favored.

\bibliography{bibliography,refs}